# Life as a Categorical Information-Handling System: An Evolutionary Information-Theoretic Model of the Holobiont

**Antonio Carvajal-Rodríguez** [1,*]

[1] Centro de Investigación Mariña (CIM), Departamento de Bioquímica, Genética e Inmunología. Universidade de Vigo, 36310 Vigo, Spain; acraaj@uvigo.es

* Correspondence: acraaj@uvigo.es

## ABSTRACT

Living systems can be understood as organized entities that capture, transform, and reproduce information. Classical gene-centered models explain adaptation through frequency changes driven by differential fitness, yet they often overlook the higher-order organization and causal closure that characterize living systems. Here we revisit several evolutionary frameworks, from the replicator equation to group selection and holobiont dynamics, and show that evolutionary change in population frequencies can be expressed as a Jeffreys divergence. Building on this foundation, we introduce a categorical model of Information Handlers (IH), entities capable of self-maintenance, mutation, and combination. This abstract architecture illustrates the usefulness of category theory for framing evolutionary processes that range from very simple to highly complex. The same categorical scheme can represent basic allele-frequency change as well as more elaborate scenarios involving reproductive interactions, symbiosis, and other organizational layers. A key feature of the framework is that different levels of evolutionary change can be summarized through a measure that quantifies the information generated, thereby distinguishing diverse types of evolutionary transformation such as individual and sexual selection, mate choice, or even holobiont selection. Finally, we show that the informational partition associated with host-microbiome pairings in holobionts generalizes the information-theoretic structure previously developed for non-random mating, revealing a common underlying architecture across biological scales.

## 1. INTRODUCCIÓN

What is life? Numerous definitions have been proposed, approached from biochemical, metabolic, thermodynamic, and other perspectives [1–5]. At the most basic level, life is associated with self-organized molecular structures that remain open to exchanges of energy and matter and are capable of self-maintenance, renewal, and reproduction. These self-organized structures are what we call organisms. Life, in turn, is the set of activities an organism performs to maintain its structure and function.

In this work, we argue that life can be understood as a hierarchical system of information handling, in which reductionist and holistic perspectives are not mutually exclusive but complementary for understanding biological complexity. Our aims are twofold: first, to show that across different hierarchical levels, the effects of information handling can be summarized by a measure of the information generated; and second, to develop an abstract formal framework that enables the description of different levels of evolutionary interaction and their informational signatures in the form of a Jeffreys divergence. Throughout this work, "information" refers to any heritable, transmissible, or associational feature—genetic, epigenetic, phenotypic, ecological, or relational—that contributes to systematic changes in population-level distributions across evolutionary steps, regardless of whether those changes are driven by selection, transmission, migration, or other processes

When life unfolds within a particular organism, we say that the organism is alive. A tree, for example, is in a living state because it performs processes such as photosynthesis and growth. Yet the tree is not isolated; it interacts with organisms of other species as well as with other trees. Fungi growing within and around its roots form a symbiosis known as mycorrhiza. Mycorrhizae are essential for normal tree development because they absorb soil nutrients inaccessible to the roots and transfer them to the tree, which in turn provides sugars required for fungal growth. As fungal networks extend through the soil, they may temporarily connect the roots of neighboring trees, forming a common mycorrhizal network (CMN) [6,7]. Thus, interactions occur among conspecifics (tree-tree) and across very different species (tree-fungus).

Living beings undoubtedly depend on other living beings for survival. Animals rely on plants for oxygen and food; plants rely on animals and microbes for pollination and for the decomposition of organic matter. Matter within the biosphere is continuously recycled: decomposers such as bacteria and fungi break down organic material and return nutrients to the ecosystem. The sun's energy, once captured by organisms through photosynthesis, flows through a complex network of interactions. This intricate web operates across multiple levels, forming what can be viewed as a hierarchy of organisms within organisms, life within life. Viruses within a cell satisfy many criteria for life; the cell within an organism is alive; bacteria within plant, animal, or fungal hosts are alive; and the organism containing them is also alive. All known life exists within what we call the biosphere. The biosphere is therefore permeated by life, and modern Earth system science suggests that it can even be viewed as a complex adaptive system capable of self-regulating the conditions necessary for life [8,9]. Life results from a network of interactions among entities that operate not only at the same level (as in cells within tissues or organisms within ecosystems), but also across levels: downward to organisms they contain, and upward to organisms that contain them. From this perspective, biological organization appears not only as something to be explained (explanandum), but as something that explains (explanans) [10,11] .

Life cannot be understood without evolution, the process that shapes biological diversity. Living beings constantly interact with their environment and with one another, driving evolutionary change. Some traits vary over time and persist either by contingency or because they enhance reproductive success. Evolution is thus the change in the ensemble of entities across generations. The capacity to evolve is a key feature distinguishing living from non-living systems. Viruses, for instance, lack autonomous metabolism but do evolve, illustrating this boundary condition. This raises a central question: how does evolution occur?

Changes may occur in the environment, the organism, or both. Environmental changes: catastrophes, accidents, or fluctuations; may cause some individuals to die while others survive. Organismal changes may produce advantages that persist only if the modifications are inherited. Although non-genetic forms of inheritance are now recognized as evolutionarily relevant, these too rest on genetic mechanisms that evolved through natural selection [12,13]. Ultimately, nucleic acids, DNA and RNA, and the information encoded in the genetic code underpin the capacity for heredity and organismal development.

This introduces a second axis of biological understanding. We began by emphasizing the relational, multilayered web of interactions that constitutes the biosphere, yet across billions of years, from the first cells, bacteria, or viruses some 3-4 billion years ago [14,15], a continuous thread can be traced: nucleic acids. They are universal to all known life and encode the information required for organisms to develop in space and time. Whatever the scenario proposed for the origin of life, all views acknowledge the centrality of nucleic acids as carriers and transmitters of genetic information. Their establishment as a universal feature of life may have been gradual, accompanied by an increasing role in biological function.

We thus encounter two focal perspectives on life and evolution: a holistic or systemic, non-gene-centric one, and a reductionist one. The holistic view holds that the whole is more than the sum of its parts and cannot be fully understood by decomposing it. It emphasizes life as an integrated network of interactions. The reductionist view maintains that the system can be understood by analyzing its components, particularly genetic information, which connects present-day organisms to the earliest life forms and enables the diversity we observe today. Although these perspectives are often presented as opposites [16–24], even holistic interpretations acknowledge that life began in simple forms that evolved into greater complexity. The RNA world hypothesis [25], despite unresolved issues, [26], illustrates this point: simple self-replicating ribozymes could undergo natural selection, enabling increasing complexity. Over time, evolutionary processes produced more agency (interactions constituting the organism and enhancing survival), more interaction, and more complexity [27–29]. This complexification suggests that reductionist and holistic approaches are complementary across temporal and organizational scales. Early life was simple enough that both perspectives converged; complexity later demanded more systemic frameworks.

Thus, beginning with a reductionist analysis, centered on simple replicating systems, can be reasonable, provided one acknowledges the eventual need for systemic approaches. Indeed, reductionist insights can be invaluable for developing such frameworks. Genomic analyses, for example, illuminate major evolutionary phenomena such as speciation [30–33] and the ecological and evolutionary roles of symbiosis and symbiogenesis [34–38].

We emphasize that a key property of living systems is information handling: the ability to acquire, interpret, and use information in functionally meaningful ways to enhance survival and reproduction. Quorum sensing in bacteria, for example, enables populations to coordinate behavior through information exchange [39,40]. Information handling itself generates new information, and this can be quantified. To show this, we begin with the simplest case, the genetic replicator, and then move to organisms carrying replicators, pairing interactions, groups of organisms, and holobionts (composite organisms consisting of a host together with its associated microbial community), demonstrating how informational patterns emerge at each level when natural selection acts.

Beyond serving as a formal abstraction, the Information Handlers (IH) framework is intended as a biological tool. Each informational fold corresponds to a measurable evolutionary process, be it selection on replicators, changes in mating patterns, or host-microbiome associations. This makes the formalism directly applicable to empirical data, where observed frequency changes can be partitioned into host, microbial, and interaction components. In this sense, the categorical architecture does not replace classical evolutionary theory but complements it by revealing the causal and informational structure underlying widely used models.

This manuscript is organized as follows. Sections 2 and 3 review foundational concepts such as replicators, phenotypes, and biological information. Section 4 examines a recently published model that analyzes the dynamics of opposing forces in the tragedy of the commons, and shows how evolutionary change in that context can also be expressed in informational terms. Section 5 reviews the connection between the extended phenotype and symbiosis and shows that, in the case of holobionts, selective dynamics can likewise be captured informationally. In Section 6, we introduce a categorical framework of information handlers, where each dynamic level is captured by an informational fold that quantifies the magnitude and functional significance of evolutionary change. Section 7 applies this categorical formulation to holobionts, yielding an informational decomposition that generalizes earlier formalisms for non-random mating. Sections 8 and 9 present the discussion and conclusions. Section 10 contains three mathematical appendices: (i) the informational partition of a simple holobiont model, (ii) the categorical framework, and (iii) its extension to incorporate symbiotic interactions and to generalize the informational partition in (i).

## 2. GENES REPLICATE, PHENOTYPES INTERACT

### 2.1. Replicators

In evolutionary biology, a replicator is any entity capable of producing copies of itself, directly or indirectly, while preserving its structure (and the information that defines it) across generations, thereby persisting over time and undergoing mutation or variation. Such variation enables the action of natural selection [41]. According to Dawkins, the fundamental unit of evolution and the primary driver of life is the replicator, although other authors have questioned the viability of the simple replicator as the origin of life [42]. In practice, when we say that natural selection tends to populate the world with more efficient replicators, what occurs is that those replicators, or more precisely, their vehicles (sensu Dawkins), that better match the environmental conditions in which replication occurs, leave more descendants. If such a match persists over time, those replicators increase in frequency within the population. What constitutes a better fit with the environment depends on the level of biological organization. For early prokaryotic life, it may have involved greater catalytic efficiency or increased resistance to environmental fluctuations. In more complex organisms, fitness may depend on traits such as locomotion, predator avoidance, resource acquisition, or mate attraction. From the perspective of the replicator, however, these diverse possibilities all translate into an increased capacity to produce copies of itself. In population genetics, replicators with higher biological fitness have greater expected reproductive success. Biological fitness can thus be defined as an entity's general propensity to survive and reproduce in a given environment. This propensity reflects the expected consequences of its interaction with the environment through its phenotypic vehicle, the organism [43,44].

## 2.2. Interactors

In defining biological fitness, we have deliberately used the term entity, since we refer both to replicators and to the units that express or carry them. The fitness of a given replicator depends on its capacity to persist and multiply within the environment in which its survival machine, the organism, operates. However, organisms contain multiple replicators (genes) that may not be transmitted together; therefore, the fitness of an individual replicator is not necessarily equivalent to that of the organism that carries it. Replicators constitute units of inheritance, but selection can operate at different hierarchical levels. Within this context arises the long-standing debate on the levels of selection [45–47], which distinguishes among (i) the entities that interact with the environment (organisms or groups), (ii) the entities that replicate or reproduce (genes, organisms, or holobionts), and (iii) the entities in which observable adaptations manifest as cumulative outcomes of evolutionary change (e.g., stinging behavior in honeybees) [48,49].

The units that interact with the environment, interactors [46], are those that exhibit phenotypic variation, and whose interaction with the environment generate differences in replication or reproduction. Replicators or reproducers, by contrast, are the units of inheritance. The iterative action of interactors and replicators across generations gives rise to adaptations [49]. From this perspective, natural selection can be defined as the process by which the differential proliferation of interactors causes the differential replication of replicators (or the differential reproduction of reproducers) [48].

Information flows through every level of biological organization. The more effectively an organism can capture, process, and use relevant environmental information, whether acting through its replicators or as an interactor, the higher its effective fitness. Replicators interact with one another within a single organism, but the organisms that contain them also interact in diverse ways [50], generating symbiotic dynamics across multiple biological scales. Under such conditions, the amount of meaningful information increases, because each informational unit acquires greater functional significance as the system to which it belongs becomes more interconnected and coherent [51]. These considerations support a view increasingly emphasized in recent years: life can be understood as matter that exchanges information and generates functional meaning, and biological evolution as a continuous process of information acquisition, transformation, and optimization [52–55]. The different levels at which information and meaning arise can thus be viewed as a hierarchy of informational levels or categories.

Since the origins of neo-Darwinism, however, evolution has often been equated with changes in gene frequencies, thereby reducing the complexity of evolutionary processes to a purely genetic dimension [56–59]. Yet, as noted above, although genes are replicators, the units that interact with the environment, and thus experience selection, are the phenotypically expressed organisms and groups that vary and proliferate differentially.

Genetic variation is undoubtedly essential to evolution, but the reductionist tendency has been reinforced not only by the central role of genetic information, but also by the pursuit of mathematical formalization. Such formalization is greatly simplified when individuals are represented as combinations of a small number of replicators, that is, as simplified genotypes. However, when the effect of a gene depends on a complex network of interactions, its true effect becomes a property of the system in which that gene is embedded, together with its local environment [60–64].

## 3. INFORMATION FLOW IN REPLICATOR-INTERACTOR SYSTEMS

Information in organisms is not static. Although a gene possesses a fixed nucleotide sequence, its expression depends on context and on multiple internal and external conditions. Genes interact with other informational pathways, generating complex regulatory networks within and beyond the cell that together form a coherent whole. Biological information is indeed encoded in genes [65,66], yet these genes are expressed as proteins that participate in signaling networks and intercellular communication, as well as in interactions between the cell and its environment. External signals are received through membrane receptors, triggering signaling cascades mediated by proteins such as kinases, G-proteins, and transcription factors that ultimately regulate gene expression [53,67]. Information in biology is therefore dynamic and functionally meaningful [53,66,68,69]. Living systems acquire information about their environment and integrate it into a hierarchical network of flows and transformations, far more complex than a linear sequence of zeros and ones. Moreover, information also travels between individuals, for example through species-specific recognition signals used in mating [70].

From an evolutionary perspective, many approaches focus on how genomes and organisms store information about the environment to which they are adapted [71–75]. However, information is not accumulated solely in the genome; it can also be measured at the population level, for instance by quantifying the informational increase caused by natural selection through changes in the distribution of frequencies [76] or even through the distribution of matings [77,78]. Consequently, although adaptive information accumulates in genomes across evolutionary time, the resulting patterns can be measured at multiple levels: population, phenotypic, and genotypic [79]. At the population level, it is also possible to quantify the information associated with the different stages of the life cycle by comparing them with their corresponding null models [80]. From a quantitative standpoint, one metric for estimating adaptive information is the Kullback–Leibler divergence (KL) [81], which measures the deviation of an observed distribution from that expected under a neutral model. When the underlying dynamics follow a replicator-type structure, either strict or extended to include mutation and recombination, so that a function Φ maps frequency changes from generation $t$ to $t + 1$ [82], the change associated with log Φ corresponds to the Jeffreys divergence [76,83]. In such cases, the Jeffreys divergence is the most appropriate measure for quantifying the total amount of information generated by selection, as it is a symmetric divergence incorporating changes in both directions.

At the ecosystem level, species diversity and their distributions also constitute a quantifiable form of information, reflecting the structural and functional complexity of ecological systems [51,84–86]. Together, these molecular, cellular, population-level, and ecosystem-level perspectives illustrate that life can be understood as a hierarchical network of informational flows.

In the following sections, we show how these flows can be measured and subsequently reinterpreted as manifestations of Information Handlers (IH), whose transformations and replications generate informational folds, defined as cumulative, structured measures of informational change within the system at each hierarchical level of organization.

### 3.1. Information dynamics of replicators

Let us briefly recall the model of replicator evolution. Consider a population of replicators, and let $w_i(t)$ denote the fitness of type *i* in generation *t*, understood as its reproductive success potential or its capacity to generate copies of itself. The expected change in the frequency of type *i*, as a function of its relative fitness with respect to the population mean, is given by the discrete replicator equation [76]:

$$p_i(t+1)=p_i(t)\frac{w_i(t)}{\bar{w}(t)}, \qquad (1)$$

where $p_i(t)$ is the frequency of replicator *i* in generation *t*, and $\bar{w}(t)$ is the mean fitness of the population in generation *t*. The change in frequency due to differential fitness is thus:

$$\Delta p_i = p_i(t+1) - p_i(t) = p_i(t)(\omega_i - 1),$$

where

$$\omega_i = \frac{w_i(t)}{\bar{w}(t)}.$$

If, instead of considering the replicator *i*, we examine an associated phenotypic trait $Z_i$, the population mean value in generation *t* is

$$\bar{Z}(t)=\sum_{i=1}^{n} p_i(t) Z_i,$$

and the mean value in the next generation, assuming that only frequencies change, is

$$\bar{Z}(t+1)=\sum_{i=1}^{n} p_i(t+1) Z_i.$$

The change in the mean value of *Z* due to differential fitness can be expressed [76] as

$$\Delta \bar{Z} = Cov(\omega, Z). \qquad (2)$$

If we take the trait *Z* to be the logarithm of relative fitness, $Z_i = \log(\omega_i)$, then (2) corresponds to a divergence Φ whose functional form is

$$\phi = J(p', p) = \sum_i (p'_i - p_i) \log \frac{p'_i}{p_i}.$$

This is the Jeffreys divergence [83,87], the symmetric version of the Kullback–Leibler divergence, which sums the two possible divergences before and after selection. In this case, (2) can be rewritten as

$$\Delta \overline{\log}(\omega) = cov(\omega, \log(\omega)) = J(p', p). \quad (3)$$

Equation (3) measures the rate of information production by natural selection [76]. $J(p', p)$ quantifies how much information is gained by the fitter replicators and how much is lost by the less fit ones. An equation of the same form as (3) can also be obtained for extended replicators, for example, when including mutation or recombination; provided that there exists a function $f(\omega)$ that maps a probability distribution $p(t)$ into $p(t+1)$:

$$p_i(t+1) = p_i(t) f(\omega_i), \quad (4)$$

and allows defining $Z = \log(f(\omega))$ as an informational trait. In this way, the evolutionary dynamics can be described in terms of the Jeffreys divergence, which measures the total flow of information reallocated during the life-cycle process.

Carvajal-Rodríguez [77,78,83,88,89] applied this framework to different contexts: non-random mating, sexual selection, and haplotype evolution; showing that the deviation from the corresponding null model is naturally expressed through $J(p', p)$. In all these cases, the term $J$ represents a symmetric measure of the total change in information, capturing the net informational flow in the population.

### 3.2. Information dynamics of interactors

As we have seen, the units of selection can be located at different hierarchical levels: replicators, which transmit information, and interactors (or vehicles), which express that information and act upon the environment. This distinction helps to disambiguate the different roles biological entities play in evolution. A single entity can behave as a replicator in one context and as an interactor in another, depending on the level of organization under consideration [48]. This raises the question: can changes in the frequency distribution of a population of organisms, or of other types of interactors, be described by equations of type (1–4)? We know that, at least in the case of mating interactions, the answer is yes. Carvajal-Rodríguez [77] showed that mating distributions can be represented by an equation of the same form as (1), where the distributions correspond to the frequencies of pair formation under different schemes (random, mate choice, or intra-sexual competition). Although mating pairs are not replicators in the strict sense, since they do not copy themselves, the dynamics of pair formation described by the equation generate measurable information in terms of the divergence *J*. In this context, *J* quantifies the total information associated with the change in the distribution of matings, such that when mate choice deviates from randomness, detectable information is generated both in the mating distribution and, potentially, in the general population (when sexual selection is present).

In a similar way, interactors can also be defined at higher levels: groups, symbiotic associations, or ecological communities. The question that then arises is: what kind of information is generated at the population or supra-individual level in these cases? Such questions motivate the next two sections, which show how changes in frequency distributions within or between groups, or in holobionts, can be described by equations of type (1-4) and how the associated informational changes can be quantified through the Jeffreys divergence *J*.

## 4. INTRA- AND INTER-GROUP SELECTION INFORMATION

In the context of group selection, natural selection can act at different hierarchical levels that often exert forces in opposing direction [47,90]. Frank [91] develops a simple tragedy of the commons model [92] to describe individual competition at the local level and group competition at the global level between two types of haploid genotypes with competitiveness values $x_1$ and $x_2$, occurring at frequencies $q$ and $1 - q$, respectively.

### 4.1. Intra-group information

The mean competitiveness within a group is

$$y = q x_1 + (1 - q) x_2,$$

and the relative competitiveness of genotype 1 is

$$\omega_1 = \frac{x_1}{y}.$$

Its frequency within the group after intragroup competition is

$$q' = q\,\omega_1.$$

The change in within-group frequency before between-group selection is

$$\Delta q = q' - q = q\,\omega_1 - q = q(\omega_1 - 1),$$

and the average change in an associated character *Z* can be written in the form of an equation of type (2). If we take $Z = \log(\omega)$, we obtain the within-group information

$$J_q = (q' - q)\log\frac{q'}{q} + (q - q')\log\frac{1-q'}{1-q},$$

which represents information generated by selection within a particular group.

The average intra-group information before inter-group selection will be

$$\bar{J}_x = \sum_q p_q J_q,$$

where $p_q$ is the probability that the group has an initial frequency *q*.

### 4.2. Inter-group information

The relative competitiveness of the entire group (normalized by the average group fitness) with a relative intensity *s* of inter-group selection is [eq. 11 in 91]

$$\omega_G = \frac{(k - y)^s}{\bar{w}},$$

where $k > 0$ is a term that avoids negative values in $\omega_G$.

The probability that a group with genotype 1 at frequency *q* contributes to the set of groups of the next generation is proportional to

$$p'_q = p_q\,\omega_G.$$

Information on the change in group distribution caused by inter-group selection, taking $Z = \log(\omega_G)$, is

$$J_G = \sum_q (p'_q - p_q)\log\frac{p'_q}{p_q}.$$

Thus, for a simple tragedy of the commons model, it is possible to quantify the flow of information associated with intra- and inter-group selection based on changes in frequency distributions. However, biological systems exist within complex networks of interactions where, in addition to competition between groups, individuals of different species are involved, and their cooperation may be essential for survival. This is particularly evident in symbiotic processes, which raises the question of what kind of information is generated through the combination of organisms engaged in symbiotic associations.

## 5. EXTENDED PHENOTYPE AND SYMBIOSIS

In *The Extended Phenotype* [50], Dawkins proposed that the effects of a gene, that is, of replicators, are not confined to the phenotype of their survival machine [the individual or interactor sensu 46], but can extend beyond it, significantly influencing the environment. The nests built by birds, the dams of beavers, or the webs of spiders can be considered expressions of these animals' genes, since such constructions and behaviors directly affect the organism's survival and reproduction.

Natural selection acts upon these extended phenotypes in the same way it acts upon traditional phenotypes: genetic variants responsible for extended phenotypes that produce favorable effects for the individuals carrying them will tend to be positively selected. Extended phenotypes associated with a gene or set of genes, such as nests, burrows, or river dams, generate effects that transcend the individual, modifying the dynamics of entire groups. Thus, the extended phenotype can be interpreted as a mechanism through which the effects of genes at one level (the individual) influence selection at another level (the group). Dawkins also noted that mutually beneficial replicators tend to predominate over others. From his gene-centered perspective, this explains the evolution of cooperative vehicles that carry replicators with complementary effects. However, when an organism, through a specific genetic variant, modifies the environment or the behavior of another organism in such a way that both benefit, the result is functionally analogous to a mutualistic symbiosis. A well-known example involves intestinal bacteria that help digest food in exchange for a stable habitat [93]; bacterial genes that favor this relationship spread because they increase their own survival and reproduction [94].

While Dawkins offered a replicator-centered explanation of cooperation, Margulis [95,96] approached symbiosis from an ecological and organizational perspective, showing how such interactions can give rise to new levels of biological complexity. From this standpoint, symbiosis is not merely an adaptive strategy but also a process that generates organization and facilitates the exchange of information between systems. When we refer specifically to the symbiosis between a host and its microbial communities, we refer to holobionts, that is, integrated systems that constitute new functional and evolutionary units.

### 5.1. Holobionts

A holobiont is a composite organism that includes a host together with the ecological community of microorganisms that constitute its microbiome [97,98]. The combined set of genomes of the host and its associated microorganisms forms the hologenome [99,100]. The hologenotype represents the particular configuration of the hologenome in an individual holobiont, and its phenotypic expression, which determines the behavior, physiology, and morphology of the system, defines the holophenotype. Holobiont selection is therefore understood as the differential reproduction or survival of holobionts according to their holophenotypes [98]. Over time, this selection modifies both the total number of holobionts and the frequency distribution of hologenotypes.

A current debate concerns whether selection acts on the host–microbiome combination as a single functional unit, or whether the two coevolve partially independently, given that microbiome transmission is not always vertical [98 and references therein]. In many cases, hologenome integration is incomplete and microbiome transmission is horizontal, which makes it difficult to regard the holobiont as a fully autonomous unit of selection. Nevertheless, a model proposed by Roughgarden [98], shows theoretically that holobiont selection remains an evolutionarily plausible force even under horizontal transmission. The model includes three stages within each generation: (1) microbial transfer among holobionts, (2) internal microbial proliferation, and (3) holobiont selection.

Under horizontal transmission, the first step is replaced by transfer from external microbial pools.

If $H''(t, n)$ is the relative frequency of holobionts containing $n$ microbes after microbial proliferation, and $W(n)$ is the fitness of a holobiont with $n$ microbes, the frequency of holobionts with $n$ microbes after reproduction is expressed as [c.f. equation 9 in 98]

$$H(t+1,n)=\frac{H''(t,n)W(n)}{\sum_n H''(t,n)W(n)}.$$

This equation predicts the frequency of the different classes of holobionts at the beginning of the next generation and constitutes a replicator-type equation. Within this framework, the rate of information production due to holobiont selection corresponds to the Jeffreys divergence $J$ between the distributions before and after reproduction.

Using a different approach, [101] apply a niche-construction model from quantitative genetics, defining a host trait $Z$ determined additively by a genetic component $G$ and a microbial component $M$ ($Z = G + M$). If $W$ denotes the fitness of the host, the effect of selection on the trait is expressed as

$$cov(W,Z)=cov(W,G+M)=cov(W,G)+cov(W,M),$$

so that selection acting on $Z$ results in indirect selection on both the host genotype and its microbiome [101].

If we take $Z = \log(W)$ and assume that fitness decomposes into three multiplicative components, $W_G$ associated with the host's genetic component, $W_M$ associated with the microbial component, and an interaction term $a_{GM}$ defining the affinity or association between $G$ and $M$:

$$W=W_G W_M a_{GM},\ \ logW=logW_G+\log W_M+\log a_{GM},$$

then the Jeffreys divergence associated with the mean informational change in $\log(W)$ can be expressed (see Mathematical Appendix 10.1 for details) as

$$J=J_G+J_M+I_{assoc}, \qquad (5)$$

where $J_G$ and $J_M$ measure the total information associated with the frequency changes mediated by selection on the genome and on the microbiome, respectively, and $I_{assoc}$ measures how much information the system uses to pair particular genotypes with particular microbiotas.

The term $I_{assoc}$ may take slightly negative values, indicating that changes in the host and microbiome marginals and changes in the non-marginal association proceed in opposite directions. In such cases, marginal reorganization dominates and erodes the specific dependence between types, reflecting a decoupling between demographic change and host–microbiome coupling. In Section 7 we show that $I_{assoc}$ decomposes into a non-negative informational association component and a non-multiplicative interaction term, the latter being responsible for the occasional negative values.

These results connect directly with Section 6, where we formalize the informational dynamics of Information Handlers (IH) within categorical hierarchies of replication and interaction. Finally, in Section 7 we show that this framework naturally extends to holobionts, where a host IH interacts with a structured microbiome, yielding a composite informational partition that generalizes earlier information-theoretic models such as non-random mating.

## 6. EVOLUTION AS CATEGORIES OF INFORMATION

Life exhibits an intricate interplay of interconnected yet differentiated processes: genes, individuals, groups, symbionts, and environments interact through dynamic networks. This motivates a perspective grounded in relational biology, which seeks to understand living systems not primarily through their material components but through the organizational and causal relations that sustain them [102,103]. Rosen formalized this idea using category theory, representing organisms as ($M$, $R$) systems in which metabolic functions and their repair are linked in a causally closed organization [104–106].

The motivation for introducing a categorical framework is therefore biological rather than mathematical. Classical evolutionary models excel at describing how frequencies change, but they do not specify how different processes such as replication, mating, transmission (including mutation, recombination, and migration) or symbiosis fit together within a single coherent architecture. The Information Handlers (IH) model provides this missing connection. It identifies the minimal operations that living systems must perform, namely copying, transforming, and combining information, and formalizes them as morphisms (arrows) whose informational effects can be quantified. This allows biologists to compare different evolutionary scenarios, including individual selection, sexual selection, group selection, or holobiont dynamics, within a unified language.

Relational biology emphasizes that life is defined not only by its material components but by the functional organization linking them [107]. This perspective connects with systemic frameworks such as autopoiesis [108,109], organizational closure [110], and biosemiotics [111], where autonomy arises from coherent functional relations rather than physical isolation. In such views, information is understood relationally, as the preservation and transformation of structure, rather than as a transmitted object.

Although organizational closure concerns functional relations, it still presupposes a quantifiable informational substrate. The multilevel information paradigm proposes that biological systems generate and interpret information across hierarchical levels, from molecules to ecosystems, each with its own domain of meaning and function [53,112]. This relational and hierarchical view provides the conceptual basis for formalizing evolution as a categorical network of information: the Information Handler (IH) model.

### 6.1. Categorical IH Model: Information Handlers with (*M*, *R*) Closure

What follows outlines a categorical model applied to a genetic-type population dynamics framework, such as an RNA-like world of self-replicating entities with minimal functional structure. Several previous works have explored the use of category theory in genetics and evolution. For example, Tuyéras [113] introduces categorical descriptions of DNA sequences, linkage, and recombination, extending these ideas to haplotype inference and population stratification [114]. Other authors have used categorical tools to formalize the structure of phylogenetic trees [115], to represent evolutionary dynamics such as Moran processes [116] or to provide structured views of genomes through "gene categories" and their morphisms [117]. These developments aim to capture biological detail at a high level of mathematical precision, and the resulting formalisms are necessarily extensive and technically rich.

The model presented here is intentionally much simpler. It uses only the minimal categorical structure required, without attempting to encode the mechanistic complexity of genetics. Our goal is not to develop advanced categorical machinery but to introduce a lightweight conceptual scaffold capable of representing the main evolutionary processes, replication, variation, and interaction; and the informational changes associated with them. Within this framework, an abstract entity called the Information Handler (IH) is defined, endowed with the capacities to replicate, mutate, and combine information across generations.

Intuitively, an Information Handler (IH) is any biological or pre-biological entity capable of capturing, transforming, and transmitting information in a way that affects its persistence and multiplication across generations. In this sense, genes, cells, organisms, populations, or composite systems such as holobionts can all be regarded as IHs at different levels of biological organization. The IH concept is therefore not tied to a specific mechanistic substrate, but to a functional role: the handling of information relevant for evolutionary dynamics. While an IH refers to a biological or pre-biological entity, the IH framework models its evolutionary behavior at the population level, through changes in the distribution of its variants over time.

The purpose of the IH model is to provide a flexible bridge between classical reductionist approaches (focused on genes, types, and fitness) and more systemic perspectives concerned with interactions and organization. IH entities represent minimal informational processors that can copy themselves, generate new variants, and reproduce in combination with others. This yields a simple but expressive representation of evolutionary dynamics, reminiscent of replicator-type systems but formulated in categorical and informational terms.

A central aspect of the model is that evolutionary processes have associated informational consequences. Each transformation such as selection, transmission, or reproduction, induces a change in the distribution of types that can be quantified as an informational fold using the Jeffreys divergence. These folds provide a coherent way to track how information is gained, lost, or restructured at successive steps of the evolutionary process.

A summary of the model's key elements, its categorical organization, and its informational interpretation is presented below. The complete formal model is provided in Mathematical Appendix 10.2.

### *6.1.1. Categorical Structure*

A category consists of objects and morphisms (arrows representing transformations). An initial object *X*(0) is defined. From *X*(0), variants *A*(1), *A*(2),..., *A*(*t*); *B*(1), *B*(2),..., *B*(*t*);... are generated through replication or reproduction (with or without mutation), where *t* denotes discrete generations from the initial configuration. For each lineage α ∈ {*A*, *B*, *C*,...}, the classes $\alpha_0(t)$, $\alpha_1(t)$,..., $\alpha_s(t)$, coexist at each step *t*, representing the accumulated substates of the lineage. Operationally, each lineage $R_\alpha$ is treated as a small category whose objects are the possible states of the lineage and whose elementary morphisms $f_{\alpha,t}$ describe discrete temporal transitions from step *t* to *t*+1.

*Objects*

The objects of $R_\alpha$ correspond to the possible states of the lineage α:

$$Ob(R_\alpha)=\{\alpha(0),\alpha(1),\ldots,\alpha(t)\}.$$

Formally, the IH framework represents the population-level states of an Information Handler as objects, and the evolutionary transitions between those states as morphisms.

Here, α(*t*) denotes the state of the lineage at step *t*. Each state may contain several substates (or variants) that coexist at time *t*. We denote these by

$$\alpha(t)=\{\alpha_0(t),\alpha_1(t),\ldots,\alpha_s(t)\},$$

where $\alpha_0(t)$ corresponds to the ancestral subtype still present at time *t*, and the index *s* labels the accumulated variants or mutational substates of the lineage. Thus, substates coexist within the same generational step, even though they originate from previous transitions.

*Morphisms (arrows)*

Morphisms describe the operations of replication, variation, or transformation that lead from one state to the next:

$$f_{\alpha,t}:\alpha(t)\rightarrow\alpha(t+1),$$

where α(*t*) is the Information Handler at step *t*, and $f_{\alpha,t}$ is the combined action of replication, with or without mutation, producing the new handler α(*t*+1).

At this level, it is not necessary to specify internal subtypes, since the arrow refers to the entire set or distribution of the lineage at time *t*. As we will see, each internal state *s* of the handler $\alpha_s(t)$ is associated with a replication rate $r_{\alpha s}(t)$.

The morphism $f_{\alpha,t}$ represents the discrete temporal transition from step *t* to *t*+1. It may or may not involve a change in the internal substate *s* depending whether replication occurs without mutation ($s' = s$), preserving the internal configuration, or with mutation or variation ($s' \neq s$), producing a new subtype within the lineage. This morphism thus captures the causal process by which the system generates its next configuration, either identical or modified, maintaining lineage continuity while potentially introducing informational change.

*Closure*

Each lineage category possesses an internal (*M*, *R*) closure:

$$(M, R): \alpha(t) \xrightarrow{\rho_t} [\alpha(t), \alpha(t+1)] \xrightarrow{ev_t} \alpha(t+1),$$

such that

$$ev_t \circ (\rho_t \times id_{\alpha(t)}) = f_t,$$

where $\rho_t$ assigns to each state the mechanism capable of reproducing it in the future, an abstraction of the production of enzymes or functions enabling replication, and $ev_t$ applies the internally generated transformations to perform the transition to the next state (see Appendix 10.2.3.1).

*Supercategory R*

The collection of lineages forms a category *R*, whose objects are the categories $R_\alpha$ and whose morphisms represent processes of reproduction or interaction between lineages (e.g., hybridization, gene transfer, symbiosis, etc.):

$$R = \{R_\alpha\}_{\alpha \in \Lambda},$$

$$Mor(R) = \{F_{\alpha \to \beta} : R_{\alpha \to \beta}\}.$$

Accordingly, the IH framework operates at least across three categorical levels: substates or types, population-level lineage states indexed in time, and lineages themselves as interacting evolutionary systems.

### 6.1.2. Discrete Dynamics within a Lineage

Each object at substate *s* in lineage α is connected to the set of copies it contains i.e. it has a population of instances

$$P_{\alpha}(\alpha(s);t) \subseteq \mathbf{Set},$$

with cardinality $N_{\alpha s}(t)$ and the frequency of subtype *s* at time *t* is

$$p_{\alpha,s}(t)=\frac{N_{\alpha,s}(t)}{\sum_{u} N_{\alpha,u}(t)}.$$

The evolution of frequencies follows the replicator equation:

$$p_{\alpha,s}(t+1)=\frac{p_{\alpha,s}(t)\cdot r_{\alpha,s}(t)}{\sum_{u} p_{\alpha,u}(t)\cdot r_{\alpha,u}(t)}, \qquad (6)$$

where $r_{\alpha,s}(t)$ is the effective replication rate including both the effects of selection and transmission. However, it is also possible to consider the strict replication rate to measure frequency change due solely to selection (see Sections 10.2.3–10.2.4 of the Mathematical Appendix). This is the same discrete replicator equation introduced in Section 3, but now with lineage (α) and subtype (*s*) indices.

Within each generation *t*, older (persistent) and new states coexist simultaneously. The change in the frequency distribution from *t* to *t*+1, according to equation (6), allows us to define a replicative informational fold measured by the Jeffreys divergence (see below):

$$Info_{\alpha}^{(rt)}(f_{\alpha,t})=J\left(p_{\alpha}(t+1),p_{\alpha}(t)\right).$$

### 6.1.3. Reproduction

In addition to replication within lineages, the model allows for reproduction or recombination between IH entities (Figure 1). The reproductive morphisms are represented as pairings modeled by a binary operation (reproductive bifunctor) that takes two lineages and produces joint offspring. That is, each possible pairing between a pair of IHs from lineages α and β has a mating fitness $m_{\alpha(t),\beta(t)}$, and the resulting pairing frequencies are $q'_{\alpha(t),\beta(t)} = q_{\alpha(t),\beta(t)}\ m_{\alpha(t),\beta(t)}$, which can be compared with those expected by chance $q_{\alpha(t),\beta(t)}$ to define a reproductive informational fold measured by the Jeffreys divergence:

$$J_{repr}(t)=J(q'(t),q(t))=\sum_{i,j\in\alpha,\beta}(q'_{ij}(t)-q_{ij}(t))\ln\frac{q'_{ij}(t)}{q_{ij}(t)}.$$

The mating distribution can be decomposed into intra-lineage (α = β) and inter-lineage (α ≠ β) components, and the corresponding informational folds can be calculated. This allows, for instance, modeling and analyzing sexual isolation between lineages, negative assortative mating, and similar processes (see Section 10.2.6 of the Mathematical Appendix for details).

After mating, the effective replication rates are determined by a natural transformation that combines rates:

$$\phi_{\alpha,\beta}:(r_{\alpha}\times r_{\beta})\rightarrow(r^{post}_{\alpha\beta}\circ\widehat{m}_{\alpha,\beta}),$$

whose components define the aggregating rule

$$r^{post}_{ij}(t)=G(r_{\alpha,i}(t),r_{\beta,j}(t)),$$

which determines the fecundity associated with the cross $i \times j$. Different choices of the function $G$ correspond to different hypotheses about fecundity following mating. For example:

$$G_{L}(x,y)=x,\ G_{R}(x,y)=y,\ G_{mean}(x,y)=(x+y)/2,$$

$$G_{geom}(x,y)=\sqrt{xy},\ G_{\lambda}(x,y)=\lambda x+(1-\lambda),$$

$$G_{hibrid}(x,y)=I_{xy}(x+y)/2,$$

where $I_{xy} \in [0,+\infty)$ is a factor representing sexual isolation, where $I_{xy} = 0$ corresponds to complete isolation.

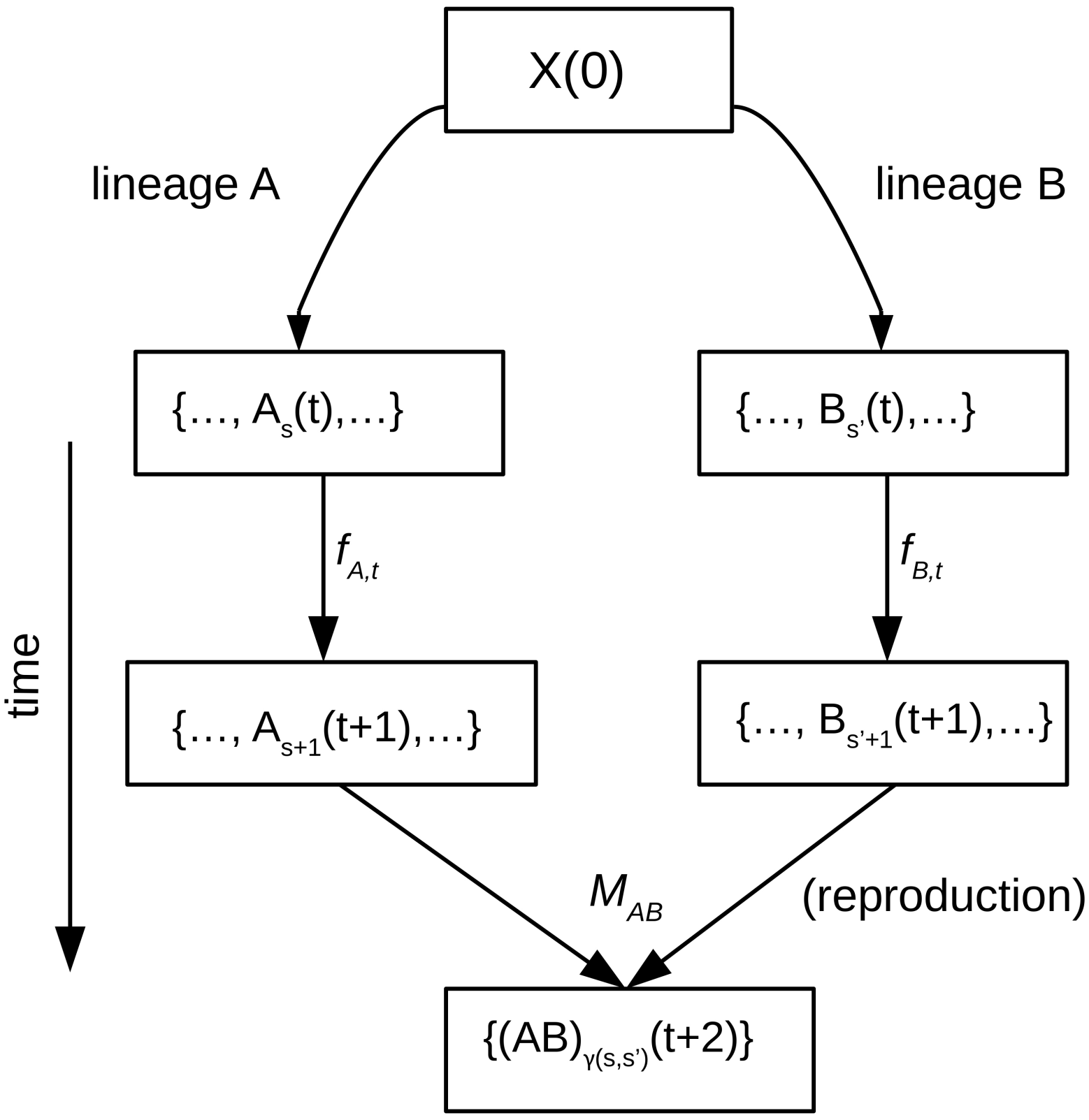

Figure 1. Basic directed graph of the IH model. The initial state *X*(0) gives rise to lineages *A* and *B*, which evolve in discrete time steps *t*; the arrows indicate the intra- and inter-lineage transformations.

#### *6.1.4. Information Functor and Biological Interpretation*

In category theory, a functor is a rule that maps one category into another while preserving its structural features. Each object in the source category (for example, a mutational state or a population) is assigned to an object in the target category (such as its statistical or informational representation), and each arrow between objects in the source category (e.g., a replication step) is mapped to an arrow between the corresponding objects in the target category (the associated informational change).

In the IH model, we work with two interconnected levels: (1) a biological level, where material transformations occur, replication, mutation, recombination and reproduction, so that Information Handlers (IHs) combine, copy, and change; and (2) an informational level, where the effects of those transformations are quantified, for instance, changes in the distribution of types within a lineage, the amount of new information generated, or the informational contribution of intra- and inter-lineage pairing events.

The information functor serves as a bridge between these two levels. It translates the biological processes by which IHs reproduce, mutate, and recombine into transformations within an informational category, where they appear as changes in frequencies and, ultimately, as changes in the system's informational content.

The informational change within each lineage is quantified by the following information functor:

$$Info_{\alpha}(f_{\alpha,t})=J(p_{\alpha}(t+1),p_{\alpha}(t)),$$

where $J(p_{\alpha}(t+1), p_{\alpha}(t))$ is the Jeffreys divergence, which measures the amount of information gained when moving between the distributions $p_{\alpha}(t)$ and $p_{\alpha}(t+1)$. Two variants are defined, $J_r$ which measures the informational change caused by selection, and $J_{rt}$ which measures the change caused by selection plus transmission. Biologically, $J_r$ quantifies how much of the informational gain is due to selection acting on the differential fitness of existing types, while the difference $J_{rt}$ - $J_r$ measures the portion attributable to non-strictly selective processes that redistribute information without conferring an immediate fitness advantage (see details of this partition in Section 10.2.5 of the Mathematical Appendix).

It is important to note that the information functor assigns an informational measure to each individual biological morphism, rather than to their composition as a single transformation. The Jeffreys divergence thus quantifies the informational change associated with each step of the evolutionary dynamics, independently of whether successive morphisms are composed. As a result, the framework does not require the Jeffreys divergence to be invariant under composition.

### *6.1.5. Total Informational Fold*

The fold, or total informational summary, generated by the system at step *t* represents the accumulated quantity of informational change due to replication, transmission and reproduction:

$$Info_{Total}(t)=J_{repr}(q'(t),q(t))+\sum_{\alpha\in\Lambda}J_{rt}(p_{\alpha}(t+1),p_{\alpha}(t)).$$

This total information fold functions as an informational closure, in the sense that it ensures each step of the biological dynamics has a corresponding informational counterpart.

The total information functor $Info_{Total}$ assigns to each biological morphism $f_{\alpha,t}$ its corresponding informational measure $Info_{Total}(t)$, quantified by the divergences $J$. The global accumulated information over a discrete interval $[0,T)$ is then obtained as the sum of the information measures assigned by the functor to each morphism over time:

$$Info_{Global}^{Total}(T)=\sum_{t=0}^{T-1} Info_{Total}(t).$$

## 6.2. Instantiations of the IH framework in classical evolutionary models

To clarify the relationship between the Information Handler (IH) framework and existing evolutionary models, it is useful to illustrate how the general categorical structure can be instantiated in well-known population genetics settings. The goal of this section is not to rederive classical results, but to show that standard evolutionary models can be naturally accommodated within the IH formalism once the population-state space and the corresponding evolutionary transformations are specified.

### *6.2.1. Haploid selection within a single lineage*

In the simplest case, consider a haploid population evolving within a single lineage. The population state at time $t$ is represented by a frequency distribution

$$\alpha(t)=q_1(t),\ldots,q_n(t),$$

where $q_i(t)$ denotes the frequency of allele $i$. Selection and other evolutionary forces define a transformation

$$q(t+1)=\varphi(q(t)),$$

which maps the population state to the next generation.

Within the IH framework, this transformation correspons to a morphism between the population states. The associated informational fold quantifies the evolutionary change induced by this progress through the Jeffreys divergence between $q(t)$ and $q(t+1)$.

In classical haploid models, evolutionary equilibria are defined as fixed points of the population mapping, $q^*=\varphi(q^*)$. Within the IH framework, such equilibria correspond to transformations that generate no informational change: the associated informational fold vanishes, $J(q^*, q^*) = 0$. Thus, equilibrium states are naturally recovered as informationally neutral configurations, while the approach to equilibrium is characterized by positive informational change quantifying the strength and direction of the evolutionary process.

### *6.2.2. Extensión to diploid populations*

Diploid evolution can be incorporated into the IH framework by redefining the population-state space to track genotype frequencies rather than allele frequencies. In this case, the population state at time *t* is represented as

$$\alpha(t)=(q_{AA}(t),q_{Aa}(t),q_{aa}(t)),$$

where each component denotes the frequency of a diploid genotype.

Reproduction, segregation, and selection define a transformation between these genotype-frequency distributions, mapping $\alpha(t)$ to $\alpha(t+1)$. From the categorical perspective, this change corresponds to an internal reconfiguration of genotype frequencies within the same population-state object, while leaving the overall IH scaffold unchanged. Evolutionary change is still represented as a transformation between distributions, and its informational effect is captured by the corresponding informational fold.

### *6.2.3. Multiple populations and migration*

The IH framework can also accommodate classical models of migration by allowing population states, corresponding to distinct subpopulations, to be connected through additional morphisms. A simple and well-known example is the island-continent model, where the island population exchanges migrants with a much larger continent whose allele frequencies remain effectively constant. In this case, the population state of the island at time *t* is represented by a frequency distribution

$$\alpha_I(t)=q_I(t)=(p(t),1-p(t)),$$

where the continent is characterized by a fixed distribution

$$\alpha_C=q_C=(P,1-P).$$

Migration acts as a transformation that mixes both distributions,

$$q_I(t+1)=(1-m)q_I(t)+mq_C,$$

where $m$ is the migration rate. This transformation defines a morphism acting on the island population state, with the continent entering as a fixed contextual parameter.

Defining the associated mapping

$$\varphi_{mig}^{qc}(q)=(1-m)q+mq_C,$$

the migration step can be written as

$$q_I(t+1)=\varphi_{mig}^{qc}(q_I(t)),$$

and the corresponding informatinal fold is given by

$$J_{mig}(t)=J(q_I(t+1),q_I(t)),$$

which quantifies the information change induced by migration.

Note that even when the evolutionary step is not multiplicative, as in migration, which is affine, the Jeffreys divergence remains well defined because it depends only on the pair of distributions ($q$, $q'$). The informational character is the component-wise log-ratio $\log(q'_i/q_i)$, and the Jeffreys divergence equals the change in the mean of this character between $q$ and $q'$. In replicator-like dynamics, this log-ratio coincides with a fitness-like multiplier, whereas in affine processes, such as the transformation induced by migration, it is interpreted as the reweighting of population frequencies between consecutive population states.

Within the IH framework, migration is thus represented as a well-defined population-state mapping, independent of selection. The equilibrium of the process corresponds to a fixed point of this morphism, $q^*_I = q_C$, at which the informational fold associated with migration vanishes. This illustrates how classical migration models can be easily accommodated as particular instantiations of the general IH structure, and how their evolutionary effects can be quantified and combined with other processes, such as selection, by composing the corresponding morphisms.

### *6.2.4. Reproduction between lineages and the transition to holobionts*

Allowing reproduction between types within a lineage, or more generally between distinct lineages, introduces a supercategory structure in which such interactions are explicitly represented. Once such reproductive or associative morphisms are permitted, composite entities can emerge as stable evolutionary units. As already outlined in Section 6.1.3, this construction also connects naturally with models of non-random mating, where reproductive interactions between different types generate systematic changes in pair frequencies. In those models, the resulting dynamics can be expressed in terms of informational measures, with Jeffreys divergence quantifying the effect of sexual selection or assortative mating [77,78].

The holobiont framework introduced in the next section can be seen as a further generalization of this idea, where association—rather than reproduction—occurs between evolutionarily distinct lineages, such as hosts and microbiomes, instead of between mating types within a single lineage. This perspective highlights that the transition to holobiont dynamics does not require introducing fundamentally new principles, but rather extends an already established informational treatment of reproductive interactions to higher levels of biological organization.

These examples illustrate that the IH framework provides a unifying categorical scaffold in which classical models can be naturally accommodated. Its novelty lies not in redefining known dynamics, but in offering a systematic informational interpretation of evolutionary change and in enabling extensions such as holobiont dynamics and multi-level selection.

## 7. HOLOBIONT CATEGORICAL MODEL

What we have developed so far is a categorical framework that provides a highly abstract and general structure capable of describing diverse models and contexts within evolutionary biology. As discussed in Section 5.1, interest in the evolutionary dynamics of holobionts has grown substantially in recent years, following a long period in which such systems were only marginally addressed in classical evolutionary models. Existing approaches include those cited in Section 5.1, as well as the works reviewed in Rodrigo [118]. Recent perspectives emphasizing the evolutionary relevance of host–microbiome interactions (e.g., [119]) further highlight the need for formal frameworks capable of analyzing holobiont-level dynamics.

Building on the categorical foundation developed above, we show that a holobiont can be represented as a focal IH interacting with a structured set of microbial IHs, leading to an informational decomposition that generalizes earlier information-theoretic models of non-random mating. This formulation reveals a common structure underlying both host-microbiome coadaptation and mate-choice dynamics. In the next section, we develop the categorical formulation of the holobiont within the IH framework. The section concludes with a toy model that serves as a proof of concept, illustrating how the holobiont informational partition can be applied.

### 7.1. Holobiont categorical representation

A holobiont is represented as a composite Information Handler built from two components: (i) a host lineage *H* (an IH belonging to some lineage α), and (ii) a microbiome *M* consisting of a family of microbial lineages and sublineages ($M_B$, $M_C$,…) that inhabit and interact with the host. The joint dynamics of *H* and *M*, as well as the information generated by selection, follow the partition expressed in equation (5) and Appendix 10.1.

Categorically, the host is represented as a lineage category $R_H$, with objects $H_i(t)$ denoting the possible host substates (e.g. genotypes), whose population-level frequencies define the host state at step *t*

$$H_i(t) \in Ob(R_H).$$

Similarly, the microbiome is represented as a monoidal composite category

$$R_M \simeq R_{MB} \otimes R_{MC} \otimes \cdots$$

whose objects $M_j(t)$ denote microbiome configurations:

$$M_j(t) \in Ob(R_M).$$

Biologically, representing the microbiome as a monoidal composite category means that the microbiome is not treated as a single evolutionary lineage, but as a structured consortium of multiple microbial lineages that coexist within the host. Each microbial group (e.g., phyla, functional guilds, or ASV-defined sublineages) retains its own identity and evolutionary dynamics, while jointly contributing to the state of the holobiont. The monoidal product formalizes this idea of co-presence without fusion: different microbial lineages evolve independently, yet their combined configuration determines the microbiome-associated component of holobiont fitness.

At the level of the supercategory $R$, we introduce a holobiont assembly bifunctor

$$Hol: R_H \times R_M \longrightarrow R_{H\times M},$$

which sends each pair of objects ($H_i(t)$,$M_j(t)$) to a holobiont object

$$H_i(t) \times M_j(t) \in Ob(R_{H\times M}),$$

and each pair of morphisms

$$f_{H,t}: H_i(t) \rightarrow H_{i'}(t+1), f_{M,t}: M_j(t) \rightarrow M_{j'}(t+1),$$

to a holobiont morphism

$$f_{H\times M,t} = Hol(f_{H,t}, f_{M,t}): H_i(t) \times M_j(t) \longrightarrow H_{i'}(t+1) \times M_{j'}(t+1).$$

Biologically, $f_{H\times M,t}$ represents one step of the combined host-microbiome dynamics (host replication and reproduction, microbial growth and transmission, and possible reassortment of microbiomes among hosts), while preserving the categorical structure inherited from $R_H$ and $R_M$.

## 7.2. Population functor and frequency dynamics

The holobiont category $R_{H\times M}$ is equipped with a population functor:

$$P_{H\times M}: R_{H\times M} \rightarrow \mathbf{Set},$$

mapping each object $H_i(t)\times M_j(t)$ to the set of holobiont instances realizing the host-microbiome pair.

The frequency of class ($i$,$j$) at time $t$ is:

$$q_{ij}(t) = \frac{N_{ij}(t)}{\sum_{i,j} N_{ij}(t)}.$$

Holobiont evolution follows a replicator-like update:

$$q'_{ij}(t) = q_{ij}(t) \cdot \frac{W_{ij}(t)}{\bar{W}_t},$$

where

$$\bar{W}_t = \sum_{i,j} q_{ij}(t) W_{ij}(t),$$

which is consistent with the general IH dynamics.

### 7.3. Decomposition of holobiont fitness

Following equation (5) and Appendix 10.1, we decompose the holobiont fitness into host, microbiome, and association components:

$$W_{ij}(t) = W_H(i,t) \cdot W_M(j,t) \cdot a_{ij}(t),$$

where $W_H$ is the host contribution, $W_M$ is the microbiome contribution and $a_{ij}$ is the host-microbiome association factor capturing non-independent interaction effects. We assume $W_H(i,t)$, $W_M(j,t)$, and $a_{ij}(t)$ are strictly positive for all $i$, $j$, $t$, so that $W_{ij}(t) > 0$ and all logarithmic terms in the informational expressions are well defined.

This type of multiplicative fitness decomposition is well established in evolutionary theory, particularly in group and multilevel selection models, where marginal contributions are naturally separated. Here, this approach is extended to holobionts by explicitly introducing an association term that captures non-independent host-microbiome effects. Importantly, this assumption does not impose interaction effects a priori: when $a_{ij} = 1$, host and microbiome contributions are independent and the association term vanishes. This decomposition is the foundation for the informational partition described below.

### 7.4. Information partition and the holobiont informational fold

The total Jeffreys divergence associated with holobiont selection is:

$$J_{H \times M} = J(q'(t), q(t)),$$

and as denoted in (5) it admits the partition

$$J_{H \times M} = J_H + J_M + I_{assoc},$$

where $J_H$ is the information generated by selection on the host, $J_M$ is the information generated by selection on the microbiome and $I_{assoc}$ is the information due to non-independent host-microbiome pairing.

The association component is:

$$I_{assoc} = \sum_{i,j} (q'_{ij} - q_{ij}) \cdot \log(a'_{ij}).$$

The association information $I_{assoc}$, encodes the information generated by non-independent pairing between host states $H_i$ and microbiome configurations $M_j$. Interestingly, this quantity admits a natural decomposition that is exactly parallel to the partition developed for non-random mating [78].

Let

$$r_{ij} := p'_i m'_j,$$

denote the product of the marginal post-selection frequencies of hosts and microbiomes, i.e. the joint distribution expected under independence once the marginal host and microbiome changes have been fixed.

Starting from the total Jeffreys divergence

$$J = \sum_{i,j} (q'_{ij} - q_{ij}) \log \frac{q'_{ij}}{q_{ij}},$$

we introduce the algebraic decompositions

$$q'_{ij} - q_{ij} = (q'_{ij} - r_{ij}) + (r_{ij} - q_{ij}),$$

$$\log \frac{q'_{ij}}{q_{ij}} = \log \frac{q'_{ij}}{r_{ij}} + \log \frac{r_{ij}}{q_{ij}}.$$

Expanding as in [78] yields four terms:

$$J = T_1 + T_2 + T_3 + T_4,$$

$$T_1 = \sum (r_{ij} - q_{ij}) \log \frac{r_{ij}}{q_{ij}},$$

$$T_2 = \sum (r_{ij} - q_{ij}) \log \frac{q'_{ij}}{r_{ij}},$$

$$T_3 = \sum (q'_{ij} - r_{ij}) \log \frac{r_{ij}}{q_{ij}},$$

$$T_4 = \sum (q'_{ij} - r_{ij}) \log \frac{q'_{ij}}{r_{ij}}.$$

With $T_3 = 0$, because $q'$ and $r$ share exactly the same marginal distributions, and the logarithmic factor decomposes additively into a host term and a microbiome term whose sums (or integrals in the continuous case) vanish when paired with the residuals of the joint distribution. The remaining terms have clear interpretations, as detailed in the subsections below.

### 7.4.1. Marginal host and microbiome contributions

The term $T_1 = J_H + J_M$ contains the information from marginal host and microbiome changes; it is the analogue of the additive sexual-selection components $J_{S1} + J_{S2}$ in the mating framework.

### 7.4.2. Holobiont "choice" (*PSI*-like) component

The quantity

$$J_{assoc} := T_4 = \sum_{i,j} (q'_{ij} - r_{ij}) \log \frac{q'_{ij}}{r_{ij}},$$

is the holobiont analogue of the pair sexual isolation term $J_{PSI}$ in non-random mating. It measures the information associated with the structure of host-microbiome pairing, given that the marginal changes in hosts and microbiomes have already been fixed.

Biologically, this corresponds to host-microbiome "choice", i.e. the emergent affinity patterns of which microbial communities become associated with which host types.

It is important to emphasize that cross-level correlations between host and microbiome are not imposed a priori in the holobiont model, but emerge naturally from the joint distribution of host-microbiome states. The term $J_{assoc}$ quantifies deviations from random host-microbiome association, and therefore captures the informational signature of non-independent pairing across levels. When appropriate statistical tests are defined, this component allows such correlations to be detected empirically rather than assumed.

Residual interaction effects that cannot be explained by association structure alone are captured by the interaction term $E_{holo}$ (Section 7.4.3). Together, $J_{assoc}$ and $E_{holo}$ provide a complete decomposition of cross-level effects, separating deviations from random association from genuinely non-additive interaction effects.

### 7.4.3. Holobiont interaction component

The term $T_2$ is

$$E_{holo} := T_2 = \sum_{i,j} (r_{ij} - q_{ij}) \log \frac{q'_{ij}}{r_{ij}},$$

which is directly parallel to the interaction residual $E$ in the earlier mating-based partitions

$J = J_{S1} + J_{S2} + J_{PSI} + E$ [77,78].

In both cases $E_{holo}$ (or $E$ ) appears only when both marginal and joint (*PSI*-like) changes occur simultaneously, it quantifies the non-additive information produced by the departure from multiplicativity of the components, and it measures the extra information required to explain the full system-level transformation once additive marginal and joint-structure components have been accounted for.

Thus,

$$I_{assoc} = J_{assoc} + E_{holo},$$

and

$$J_{H \times M} = J_H + J_M + J_{assoc} + E_{holo}. \qquad (7)$$

From equation (7), it would be possible to design statistical tests and estimators analogous to those developed in [77,89] that, using frequencies in $H$ and $M$, could test for selection acting on either component, as well as for possible associations between host and microbial genomes.

#### *7.4.4. The holobiont model as a natural generalization of information-theoretic mating models*

The sum of terms $J_H + J_M$ is directly analogous to the additive sexual-selection components $J_{S1} + J_{S2}$ in the mating framework. The quantity $J_{assoc}$ represents the pure association term, equivalent to $J_{PSI}$ in the mating model, capturing the structure of host-microbiome pairings beyond the product of their marginal frequencies. The component $E_{holo}$ is the non-additive interaction term, analogous in meaning and role to the $E$ term in previous models of non-random mating and sexual selection. In both contexts, the $E$-term measures the extra information generated by non-multiplicative combinations of two components, sexes in the mating model, and host-microbiome lineages in the holobiont model. Therefore, $J_{H \times M}$ constitutes a natural generalization of the informational partition originally defined for non-random mating [77,78].

The model can be extended by considering the microbiome as a structured assemblage of multiple microbial components (e.g., phyla, functional guilds, or ASV-defined sublineages). To capture this structure, we decompose the microbiome into $K$ components, each with its own internal dynamics. A holobiont state can thus be indexed by a host type $i$ and a vector of microbiome types $(j_1,\ldots,j_K)$, where $j_k$ denotes the state of the $k$-th microbiome component. The general model is described in detail in Appendix 10.3:

$$J=(J_H+\sum_{k=1}^{K} J_{M_k})+J_{assoc}+E_{multi},$$

where

$$J_{assoc}=\sum (q'-r)\log\frac{q'}{r}, \qquad E_{multi}:=T_2=\sum (r-q)\log\frac{q'}{r}.$$

This general model reduces to the two-component holobiont model in (7) when $K = 1$, and to the mating-based partitions when the two interacting components correspond to the sexes rather than to host and microbiome. From this perspective, the holobiont represents a particular instance of an Information Handler (IH) interacting with a structured ensemble of other IHs, while the information-theoretic partition developed for non-random mating [77,78] appears as the special case in which a focal IH (e.g., a female) interacts with a single partner IH (e.g., a male). In contrast, the general model provides its natural extension to multi-component associations, yielding a composite informational fold that integrates changes arising from the host, from the microbiome, and from their interaction.

### 7.5. Holobiont closure structure

Both the host and microbiome each possess their own $(M, R)$-closure:

$$(\rho_H, evt_H),\ (\rho_M, evt_M).$$

The holobiont closure is defined as a composed structure:

$$(\rho_{H\times M}, evt_{H\times M})=\Phi((\rho_H, evt_H),(\rho_M, evt_M), a_{ij}),$$

where $\Phi$ is a closure-composition operator mapping two internal closures together with an association rule $a_{ij}$ into a new effective closure.

This expresses the fact that a holobiont reproduces not only hosts and microbes, but also the patterns of association between them.

## 7.6. Holobiont information functor

Finally, we define the holobiont information functor:

$$Info_{H\times M}: R_{H\times M} \rightarrow Info(R_{H\times M}),$$

which assigns to each morphism $f_{H\times M,t}$ the informational triple:

$$(J_H, J_M, I_{assoc}),$$

or the total divergence $J_{H\times M,t}$ with its decomposition.

Thus, holobionts arise in the IH framework as higher-order composite Information Handlers, whose dynamics and informational closure naturally integrate into the global fold described in Appendix 10.2.7.

## 7.7. A Toy Model Illustrating the Informational Partition of the Holobiont

To illustrate the practical use of the informational decomposition developed in this section, we present a simple 2×2 holobiont model involving two host genotypes and two microbiome types. This toy model allows us to examine how the total informational change generated by selection partitions into host, microbiome, and association components, and it highlights situations in which the association term dominates the response. In particular, by selecting fitness values that introduce a strong synergistic advantage for a specific host-microbiome pairing, we obtain a scenario in which more than half of the total Jeffreys information arises from the host-microbiome association, providing a clear illustration of the biological meaning and utility of the term $J_{assoc}$.

### *7.7.1. Model specification*

Let the host take states $H = \{H_0, H_1\}$ and the microbiome $M = \{M_0, M_1\}$ and assume initial independence, $q_{ij} = p_i m_j$, with $p = (0.5, 0.5)$ and $m = (0.5, 0.5)$ so that all four host-microbiome combinations have frequency $q_{ij} = 0.25$.

Suppose that host genotype $H_1$ has intrinsic advantage, the same as microbiome type $M_1$, and, critically, the combination ($H_1$, $M_1$) yields a synergistic advantage. Let

$$W = \begin{pmatrix} 1 & 0.7 \\ 0.7 & 2 \end{pmatrix},$$

with rows corresponding to $H_0$, $H_1$ and colums to $M_0$, $M_1$.

The post-selection distribution is

$$q'_{ij} = q_{ij} \cdot \frac{W_{ij}}{\bar{W}}, \;\; \bar{W} = 1.1.$$

This yields:

$$q' = \begin{pmatrix} 0.2273 & 0.1591 \\ 0.1591 & 0.4545 \end{pmatrix}.$$

### 7.7.2. Marginal changes

The host and microbiome marginal frequencies are $p'$ = $m'$ = (0.3864, 0.6136) showing a clear shift favouring $H_1$ and $M_1$.

### 7.7.3. Informational decomposition

The total Jeffreys divergence generated by selection is

$$J_{H \times M} = \sum_{i,j} (q'_{ij} - q_{ij}) \log \frac{q'_{ij}}{q_{ij}} = 0.2066.$$

Partitioning according to the holobiont decomposition

$$J_{H \times M} = J_H + J_M + J_{assoc} + E_{holo},$$

we obtain

$$J_H = \sum_i (p'_i - p_i) \log \frac{p'_i}{p_i} = 0.0526,$$

$$J_M = \sum_j (m'_j - m_j) \log \frac{m'_j}{m_j} = 0.0526,$$

$$J_{assoc} = \sum_{i,j} (q'_{ij} - p'_i m'_j) \log \frac{q'_{ij}}{p'_i m'_j} = 0.1097,$$

and

$$E_{holo} = E - (J_H + J_M + J_{assoc}) = 0.2066 - 0.21489 = -0.0082.$$

In this parametrization, the residual interaction term $E_{holo}$ is small, reflecting a near exact additivity between the marginal contributions $J_H$ and $J_M$ and the pure association component $J_{assoc}$.

#### *7.7.4. Interpretation*

This example produces a holobiont scenario in which the association term dominates the informational response. The marginal components $J_H$ and $J_M$ contribute roughly one quarter of the total information each, reflecting modest selection acting independently on the host and the microbiome. In contrast, the association component $J_{assoc}$ accounts for more than half of the total Jeffreys divergence. This shows that the primary evolutionary signal does not lie in the separate advantages of host genotype $H_1$ or microbiome type $M_1$, but in the synergistic advantage of their joint occurrence. The system therefore evolves toward a coadapted host-microbiome configuration. Even in this minimal model, the informational fold reveals an evolutionary pattern that would remain hidden if host fitness or microbiome composition were analyzed in isolation.

Although presented as a toy model, this example illustrates the type of biological inference enabled by the IH framework. Any empirical dataset containing host genotypes and microbiome compositions can be analyzed by decomposing the total Jeffreys divergence into host, microbial, and association components. This allows researchers to quantify whether evolutionary change is primarily driven by host adaptation, microbial adaptation, or by the emergence of specific host–microbiome associations. Such a decomposition is not available in classical holobiont models and highlights the practical value of the informational partition.

## 8. DISCUSSION

In this work, we have reviewed several representative evolutionary models, from the simple replicator equation [57,120,121], to tragedy-of-the-commons dynamics [91], and recent models of holobiont selection [98,101], to illustrate that evolutionary change in frequency distributions can be expressed through the Jeffreys divergence [76]. This divergence also captures the redistribution of pairings caused by differential mating fitness [77,78]. Building on these results, we have proposed a conceptual framework grounded in category theory and centered on the notion of the Information

Handler (IH). IHs are entities capable of self-maintenance, replication, mutation, and combination with other handlers. Lineages of IHs evolve on a discrete temporal scale, with each lineage represented as an object in a hierarchical category whose morphisms describe processes such as hybridization, sexual isolation, or symbiosis. This abstraction applies equally to ribozymes, bacteria, animals, or plants.

Interactions such as reproduction usually occur between members of the same lineage, whereas holobiont associations may involve evolutionarily distant taxa. As shown in Section 7, the categorical structure naturally accommodates both cases, preserving the corresponding informational fold. Furthermore, the model in which a focal IH interacts with a structured set of IHs generalizes the information framework for non-random mating: the latter arises as a special case when the interacting entities are two IHs of different sexes. At each process: replication with or without selection or mutation, reproduction within or across lineages, or more complex interactions; an informational fold captures the change in the system's distribution of states.

From a biological perspective, this matters because it transforms frequency changes into interpretable evolutionary quantities. For instance, the IH decomposition allows one to distinguish whether a shift in holobiont composition is driven by selection acting on host genotypes, on microbial lineages, or on their association structure. Thus, the IH framework does not presuppose whether holobionts constitute units of selection in a strict sense, but instead provides a formal way to disentangle selection acting on hosts, microbial partners, and their associations, thereby addressing long-standing debates on the conditions under which holobionts can be treated as evolutionary individuals.

Similarly, in mating systems the IH fold identifies the contribution of mate choice, sexual isolation, and marginal selection. These examples show that the categorical structure is not merely a reformulation; it provides biologically meaningful partitions that clarify the causal pathways through which evolutionary change occurs.

This connection is particularly relevant in light of ongoing efforts to extend classical evolutionary theory beyond a strictly gene-centered view, as articulated, for example, in the Extended

Evolutionary Synthesis [122]. More generally, it resonates with two major traditions in contemporary biology: the neo-Darwinian, gene-centered view, and the systemic or organizational approach that emphasizes biological organization and the closure of constraints [11,24,107,123–126]. Under the latter, organisms are thermodynamically open but organizationally closed systems whose relatively stable constraints (e.g., enzymes, membranes, or vascular structures) modulate flows of matter and energy while depending on processes they themselves enable. Such networks of mutually dependent constraints provide stability and self-determination while allowing functional and evolutionary change. Information plays a central role in this context [52–55,112,127–130]. Distributions of genotypes, haplotypes, or matings contain more information when types differ in fitness, and less when they are equivalent or interactions are random. Although information has sometimes been criticized as metaphorical [131–133], its causal relevance lies in how systems interpret and transform it: syntactic structure and semantic meaning must be coherently integrated for biological function and cognition to emerge.

The use of Jeffreys divergence in the IH framework is robust to deviations from classical frequency-based or replicator dynamics. Jeffreys divergence does not depend on the specific biological mechanism generating evolutionary change, but solely on the existence of an initial and a derived distribution. Thus, Jeffreys divergence is not imposed *ad hoc*, but follows from the combination of a distributional mapping, a logarithmic characterization of change, and the requirement of an interpretable measure of evolutionary change. As long as an evolutionary process: genetic, epigenetic, developmental, or ecological; can be represented as a transformation between observable distributions, the informational interpretation in terms of $J(q',q)$ remains valid.

Importantly, the Jeffreys divergence quantifies the magnitude of distributional change, not its causal origin. Both selective processes and stochastic transmission processes, such as drift or migration, may generate non-zero values of $J$ when they induce systematic changes in population frequencies (for example, directional migration). The distinction between selection-driven and transmission-driven change therefore arises from the specification of the evolutionary process, the choice of an appropriate null model, and process-specific statistical contrasts, rather than from the divergence itself.

Processes such as epigenetic inheritance or developmental constraints do not invalidate this approach; rather, they modify the way population-level distributions are transformed. Epigenetic inheritance introduces transmissible states with their own dynamics, while developmental constraints restrict or bias the set of accessible states. In both cases, Jeffreys divergence captures the net informational change induced by these transformations, without requiring a replicator-based description. Accordingly, the IH framework does not depend on replicator dynamics in a strict sense. Replicator equations are treated as particular instances of a more general mapping φ that transforms an initial distribution $q$ into a derived distribution $q'$. Whenever such a mapping can be defined, the corresponding informational change is naturally quantified by the Jeffreys divergence.

The IH framework thus serves two purposes: it highlights how natural selection can be captured through informational changes in frequency distributions at multiple levels (individuals, pairs, groups, holobionts), and it illustrates how category theory can provide a formal scaffold for integrating the informational and dynamical aspects of biological evolution. While the categorical abstraction does not aim to preserve fine-grained molecular or physiological mechanisms, it explicitly preserves causal structure at the level relevant for evolutionary dynamics. In particular, interactions between evolutionarily distant taxa, such as hosts and microbiomes, are represented through distinct components and an explicit interaction term, allowing their causal and non-additive effects on evolutionary change to be quantified without assuming shared mechanistic substrates.

Other recent formalisms pursue similar goals. Smith [80], for example, decomposes the evolutionary process into stages treated as informational channels. The IH framework formalizes a comparable intuition but at a categorical and hierarchical level. It yields a global dynamical model in which each transformation of the system is represented as a morphism associated with an informational fold quantified by the Jeffreys divergence arising from the underlying evolutionary dynamics.

Frank [134,135], in turn, shows that the Price equation provides a universal pattern for partitioning evolutionary change and identifies an underlying geometric structure shared by natural selection,

Bayesian inference, thermodynamics, learning algorithms, and informational metrics. In his approach, measures such as the Kullback–Leibler divergence, the Jeffreys divergence, and the Fisher metric follow necessarily from structural invariances (e.g., probability conservation). The IH model is compatible with this view: each informational fold $J$ can be formulated explicitly via the Price equation, and the Jeffreys divergence associated with each transformation corresponds exactly to the change term obtained when applying Price to an appropriate logarithmic trait under replicator-like dynamics. However, the Price equation alone does not supply lineage structure, organizational closure, or hierarchical categories of entities and levels. It describes how a distribution changes between two configurations, but not a unified framework capable of encompassing classical replication, sexual selection, or holobiont evolution. That integrative architecture is provided by the categorical formalization of the IH model, within which the informational folds derived from Price are embedded as natural quantifiers of change. In this sense, the IH formalism yields a unified global model whose informational partitions are derived directly from the causal operators governing replication, transmission, and reproduction.

Within this broader theoretical perspective, life may be understood as a process of managing and transforming information, where informational closure ensures that the informational changes relevant for persistence and reproduction are generated and evaluated within the system itself. In this framework, information is not understood merely as a statistical descriptor of variation, but as a measure of functionally relevant change. While Jeffreys divergence is formally defined on probability distributions, its biological interpretation arises from the fact that these distributions describe the frequencies of states that differ in their causal and functional consequences for the system. Informational change thus reflects not only how distributions vary, but how evolutionary processes reshape the organization of functionally meaningful states.

In the IH framework, the notion of closure is not merely metaphorical, but is formally encoded through the internal (M,R) structure of each lineage, which provides a categorical representation of causal feedback without introducing additional endomorphisms beyond those already defined. “Closure” here means that informational transitions are generated through internal morphisms:

while information flows across system boundaries, its interpretation and causal relevance remain internal. Informational closure does not imply environmental isolation but semantic autonomy: living systems generate and maintain the informational transformations relevant to their organization.

The IH framework is explicitly compatible with multiple, nested levels of biological organization. Information Handlers are not restricted to a single scale: genes, cells, organisms, populations, and holobionts can all be treated as IHs at different hierarchical levels, each characterized by its own state space, evolutionary transformations, and informational dynamics. Category theory naturally accommodates such hierarchies, allowing evolutionary processes to be represented consistently across levels without privileging a single unit of selection.

In this sense, the IH framework does not prescribe a unique unit of evolution, but provides a formal setting in which different units—individuals, groups, or holobionts—can be analyzed on equal footing, in line with classic and modern discussions on units of selection (e.g., [45,47,49]). Causal closure is understood as a level-specific property: at each organizational level, evolutionary transitions are generated through internal morphisms that encode how that level produces its own future states, while remaining open to interactions with other levels.

Thermodynamically open yet informationally closed, organisms sustain encoding, interpretation, and transmission within a self-consistent causal network. This endogenous closure grants them functional and evolutionary coherence. Life, therefore, may be viewed as a causally organized flow in which structure, function, and meaning co-emerge.

Thus, the IH framework can be used in empirical evolutionary genomics: once frequency data are available across stages of the life cycle, the informational folds can be computed directly and interpreted biologically, providing insight into the mechanisms driving evolutionary change. More broadly, recent perspectives have emphasized the role of information as a unifying principle across biological and evolutionary processes, and even beyond living systems (e.g., [136]), highlighting the growing interest in information-centered views of life, to which the IH framework contributes from a formal evolutionary standpoint.

## 9. CONCLUSION

The IH framework integrates the main evolutionary processes: replication, transmission (including mutation, recombination, and migration), reproduction, and symbiotic interaction; within a single categorical model. Each process gives rise, automatically and without additional assumptions, to an informational measure expressed as a Jeffreys divergence, capturing the magnitude of the evolutionary change produced at that step and enabling its biological interpretation within the appropriate process-specific context.

Along the way, we have also developed a new informational partition for a holobiont model, showing that the total change can be decomposed into host, microbial, and association components. This partition generalizes the information-theoretic decomposition previously obtained for non-random mating, which emerges as a special case.

Together, these results show that the IH approach provides a unified and operational framework in which evolutionary transformations, and their informational consequences, can be analyzed coherently across different biological levels, including holobionts.

## 10. MATHEMATICAL APPENDIX

### 10.1. SELECTION INFORMATION MODEL IN THE HOLOBIONT

Let $p_i$ be the frequency of host type *i* and $m_j$ the frequency of microbiome type *j*. Then, before selection, the expected frequency of holobionts of type $i{\times}j$ is

$$q_{ij} = p_i m_j .$$

#### 10.1.1. Multiplicative selection with affinities

$$W_{ij} := W_G(i) \times W_M(j) \times a_{ij} \quad \forall i, j : W_G(i) > 0, W_M(j) > 0, a_{ij} > 0,$$

$$\bar{W} := \sum_{i,j} p_i m_j W_{ij},$$

$$q'_{ij} = p_i m_j \frac{W_{ij}}{\bar{W}} .$$

### 10.1.2. Marginal frequencies after selection

$$p'_i = \sum_j q'_{ij} = p_i W_G(i) \frac{A_i}{\bar{W}},$$

$$m'_j = \sum_i q'_{ij} = m_j W_M(j) \frac{B_j}{\bar{W}},$$

where

$$A_i := \sum_j m_j W_M(j) a_{ij},$$

$$B_j := \sum_i p_i W_G(i) a_{ij},$$

and the residual affinity with respect to the new marginal frequencies:

$$a'_{ij} := \frac{q'_{ij}}{p'_i m'_j} = a_{ij} \frac{\bar{W}}{A_i B_j},$$

because

$$q'_{ij} = p_i m_j \frac{W_{ij}}{\bar{W}} = p'_i \cdot m'_j \cdot a'_{ij} \;\Rightarrow\; p_i m_j \frac{W_{ij}}{\bar{W}} = \frac{p_i W_G(i) A_i}{\bar{W}} \cdot \frac{m_j W_M(j) B_j}{\bar{W}} \cdot a'_{ij} \;\Rightarrow\; a_{ij}/\bar{W} = \frac{A_i B_j}{\bar{W}^2} a'_{ij},$$

so

$$a'_{ij} = \frac{a_{ij} \cdot \bar{W}}{A_i B_j}.$$

### 10.1.3. General information and partitioning

Taking $Z = \log(W/\bar{W})$ the average change of *Z* due to selection is

$$J := \sum_{i,j} (q'_{ij} - q_{ij}) \log \frac{q'_{ij}}{q_{ij}},$$

and we obtain the partition

$$J = J_G + J_M + I_{assoc},$$

where

$$J_G = \sum_i (p'_i - p_i) \log \frac{p'_i}{p_i} = \sum_i (p'_i - p_i) \log \frac{W_G(i) A_i}{\bar{W}},$$

$$J_M = \sum_j (m'_j - m_j) \log \frac{m'_j}{m_j} = \sum_j (m'_j - m_j) \log \frac{W_M(j) B_j}{\bar{W}},$$

$$I_{assoc} := \sum_{i,j} (q'_{ij} - q_{ij}) \log(a'_{ij}).$$

### *10.1.4. Verification*

$$q_{ij} = p_i m_j, \; q'_{ij} = p'_i m'_j a'_{ij},$$

$$\log \frac{q'_{ij}}{q_{ij}} = \log \frac{p'_i}{p_i} + \log \frac{m'_j}{m_j} + \log(a'_{ij}).$$

Then we can express $J$ as

$$J = \sum_{i,j} (q'_{ij} - q_{ij}) \cdot [\log \frac{p'_i}{p_i} + \log \frac{m'_j}{m_j} + \log(a'_{ij})].$$

We separate it into three sums and note that

$$\sum_j (q'_{ij} - q_{ij}) = p'_i - p_i$$

and

$$\sum_i (q'_{ij} - q_{ij}) = m'_j - m_j,$$

we obtain

$$J_G := \sum_i (p'_i - p_i) \log \frac{p'_i}{p_i},$$

$$J_M := \sum_j (m'_j - m_j) \log \frac{m'_j}{m_j},$$

$$I_{assoc} := \sum_{i,j} (q'_{ij} - q_{ij}) \log(a'_{ij}) = \sum_{i,j} (q'_{ij} - q_{ij}) \log \frac{a_{ij} \cdot \bar{W}}{A_i \cdot B_j}.$$

Multiplicative selection moves the marginals ($J_G$, $J_M$) through $W_G$ and $W_M$, while the host-microbiota pairing structure is captured in $I_{assoc}$ via affinity $a_{ij}$ versus their row/column averages ($A_i$, $B_j$).

## 10.2. CATEGORICAL MODEL OF INFORMATION HANDLERS WITH (*M*,*R*)-CLOSURE AND INFORMATIONAL FOLDS

### 10.2.1. Quiver and Lineage Categories $R_\alpha$ (IH)

#### *10.2.1.1. Basic Quiver (Directed Graph)*

An initial object $X(0)$ is defined. From $X(0)$, variants $A(1), A(2),\ldots, A(t)$; $B(1), B(2),\ldots, B(t)$;... are generated through replication with or without mutation, where $t$ represents discrete generations from the initial configuration. For each linage $\alpha \in \{A, B, C,\ldots\}$: at each step $t$ the classes $\alpha_0(t),\alpha_1(t), \ldots,\alpha_t(t)$, coexist, representing the accumulated substates of the lineage. This step-by-step dynamic can be viewed as a discrete coalgebra, that is, a rule that, given the current state, produces the next one. Biologically, a coalgebra describes how a living system transforms or replicates, generating its future state from the present.

#### *10.2.1.2. Definition (Lineage Category $R_\alpha$)*

*Level-0 Objects (Informational Matter)*

$$Ob_0(R_\alpha)=\{X(0),\alpha(t),\ldots\}.$$

These are the concrete information handlers. $X(0)$ is the original handler from which all others derive. Each $\alpha(t)$ represents a discrete material state of the lineage (molecule, replicator, population, or configuration). The morphisms between these objects describe material transformations: replication, mutation, or reproduction.

*Level-1 Objects (Functional)*

$$Ob_1(R_\alpha)=\{[X(0),\alpha(1)]\}\cup\{[\alpha(t),\alpha(t+1)] | t\geq 1\}.$$

For each pair ($\alpha(t)$, $\alpha(t+1)$), there exists a second level object [$\alpha(t)$, $\alpha(t+1)$] representing the functional space of the possible transformations from $\alpha(t)$ in $\alpha(t+1)$:

$$[\alpha(t),\alpha(t+1)]\in Ob(R_\alpha).$$

Formally, [$\alpha(t)$, $\alpha(t+1)$] is an internal object of the category, but biologically it represents the rules, enzymes, or functions that carry out the transformation from one material state to another.

#### *10.2.1.3. Morphisms (arrows)*

Each elementary morphism of lineage $\alpha$ is denoted

$$f_{\alpha,t}: \alpha(t) \rightarrow \alpha(t+1),$$

where α(*t*) is the information handler at step *t* (informational matter), and $f_{\alpha,t}$ Is the combined action of replication and mutation producing the new IH α(*t*+1). Here, it is not necessary to specify internal subtypes *s*, since the arrow refers to the entire set or distribution of the lineage at time *t*.

Additionally, two internal arrows are defined to link the material level with the functional one:

$$\rho_t : \alpha(t) \rightarrow [\alpha(t), \alpha(t+1)], \;\; ev_t : [\alpha(t), \alpha(t+1)] \rightarrow \alpha(t+1).$$

The arrow $\rho_t$ assigns to each resulting state α(*t*+1) the mechanism that can reproduce it in the future; it is an abstraction of the production of enzymes or functions enabling replication. Its composition with $ev_t$ generates the effective transformation of the system:

$$f_{\alpha t} = ev_t \circ \rho_t.$$

Altogether, the system produces the means that ensure its own future replication, thus fulfilling the condition of discrete (*M*, *R*) closure:

$$(M, R): \alpha(t) \xrightarrow{\rho_t} [\alpha(t), \alpha(t+1)] \xrightarrow{ev_t} \alpha(t+1).$$

See the Reproduction section 10.2.6.2 for more details on the replication.

### 10.2.2. Supercategory *R* (Category of Categories)

The set of all lineages forms a supercategory *R*, whose objects are the categories $R_\alpha$ and whose morphisms are functors between them. These morphisms represent processes of reproduction or interaction between lineages (e.g., hybridization, gene transfer, symbiosis, etc.). The natural transformations between these functors, representing coherent changes, for instance, in modes or preferences of pairing, endow *R* with the structure of a bicategory, in which the composition of processes is preserved not strictly but up to natural equivalence, that is, up to a structural correspondence that maintains functional coherence among different modes of interaction.

In other words, the composition of processes is preserved in such a way that the biological and evolutionary coherence of the system is maintained.

$$Ob(R) = \{R_\alpha \mid \alpha \in \Lambda\},$$

$$Mor(R)=\{F: R_{\alpha} \rightarrow R_{\beta} | F \text{ is a functor}\}.$$

### 10.2.3. Populations: Deterministic Functor to Set

Each object at state $s$ in lineage α is connected to the set of copies it contains. To do this, for each lineage α, we define a deterministic population functor:

$$P_{\alpha}: R_{\alpha} \rightarrow \mathbf{Set},$$

which assigns to each object $\alpha_s(t)$ a set of instances of the information handler, and to each morphism a deterministic function between sets.

#### *10.2.3.1. Objects*

Each abstract object $\alpha_s(t) \in Ob(R_{\alpha})$ is realized as a set of temporal instances given by

$$P_{\alpha}(\alpha(s);t).$$

#### *10.2.3.2. Morphisms*

For each elementary morphism $f_{\alpha,s}$ in $R_{\alpha}$,

$$f_{\alpha,s}: \alpha(s) \rightarrow \alpha(s+1),$$

the functor

$$P_{\alpha}(f_{\alpha,s};t): P_{\alpha}(\alpha(s);t) \rightarrow P_{\alpha}(\alpha(s+1);t+1),$$

associates the population of subtype $\alpha_s(t)$ at time $t$ with the population of subtype $\alpha_{s+1}(t+1)$ at time $t$+1. The increment from $s$ to $s$+1 reflects the appearance of at least one new variant during replication of the subtype, so that the time step also incorporates the genetic innovation accumulated within the lineage.

#### *10.2.3.3. Functorial Property*

$$P_{\alpha}(id_{\alpha(s)})=id_{P_{\alpha}(\alpha(s);t)},$$

$$P_{\alpha}(f_{\alpha,s+1} \circ f_{\alpha,s})=P_{\alpha}(f_{\alpha,s+1}) \circ P_{\alpha}(f_{\alpha,s}).$$

#### *10.2.3.4. Cardinalities and Frequencies per Lineage*

Let

$$N_{\alpha,s}(t)=|P_{\alpha}(\alpha(s);t)|$$

denote the number of IHs of subtype $s$ that are still present at time $t$; where

$$P_\alpha(f_{\alpha,s};t)$$

denote the replication/mutation function.

Then define

$$N_\alpha^{tot}(t)=\sum_{s\le t} N_{\alpha,s}(t),$$

$$p_{\alpha,s}(t)=\frac{N_{\alpha,s}(t)}{N_\alpha^{tot}(t)},$$

$$\sum_{s\le t} p_{\alpha,s}(t)=1,$$

where

$p_{\alpha,s}(t)$ denote the current frequency of that subtype.

#### *10.2.3.5. Functional Summary*

The functor $P_\alpha$ links the categorical structure of the lineage

$$R_\alpha=\{\alpha(0),\alpha(1),\ldots,\alpha(T)\},$$

with its concrete realization in population sets

$$P_\alpha(\alpha(s);t)\subseteq\mathbf{Set},$$

allowing to define

$$N_{\alpha,s}(t+1)=r_{\alpha,s}(t)N_{\alpha,s}(t),$$

where $r_{\alpha,s}(t)$ is the effective growth or persistence rate of each type $s$, including the net effect of entries and exits due to mutation or extinction of that type.

### 10.2.4. Intra-Lineage Replicator Dynamics

Given the set of substates {α(0),α(1),...,α(t)} of lineage α at step $t$ and letting $p_{\alpha s}$ denote the current frequency of subtype $s$ at that time, the frequency in the next generation, considering only differences in the replication rate (i.e., with no mutation or other transmission effects), is given by

$$p_{\alpha,s}(t+1)=\frac{p_{\alpha,s}(t)\cdot r_{\alpha,s}(t)}{\bar{r}_\alpha(t)}, \qquad \text{(A1)}$$

where

$$\bar{r}_{\alpha}(t)=\sum_{s} r_{\alpha,s}(t)\cdot p_{\alpha,s}(t),$$

is the mean replication rate in the population.

#### *10.2.4.1. Consistency between Counts and the Replicator Equation*

Assuming that, in the transition from *t* to *t*+1, new objects $\alpha_{t+1}(t+1)$ are generated, belonging to a new class *s* = *t*+1, and that some classes *s* ≤ *t* may disappear (i.e., become the empty set), then

$$N_{\alpha,s}(t+1)=\sum_{i} N_{\alpha,i}(t)\cdot r_{\alpha,i}(t)\cdot Q_{i\to s}(t), \qquad \text{(A2)}$$

where *Q* is a stochastic transition matrix that captures the probabilities of transition between states.

Thus we obtain

$$N_{\alpha}^{tot}(t+1)=\sum_{s} N_{\alpha,s}(t+1)=\sum_{s}\sum_{i} N_{\alpha,i}(t)\cdot r_{\alpha,i}(t)\cdot Q_{i\to s}(t)=\sum_{i} N_{\alpha,i}(t)\cdot r_{\alpha,i}(t)\cdot \sum_{s} Q_{i\to s}(t),$$

so

$$N_{\alpha}^{tot}(t+1)=\sum_{i} N_{\alpha,i}(t)\cdot r_{\alpha,i}(t)=N_{\alpha}^{tot}(t)\cdot \bar{r}_{\alpha}(t) \text{ because } \sum_{s} Q_{i\to s}(t)=1\,\forall i,$$

and dividing both sides of (A2) by $N_{\alpha}^{tot}(t+1)$ and substituting $N_{\alpha}^{tot}(t+1)=\bar{r}_{\alpha}(t)\cdot N_{\alpha}^{tot}(t)$ into the right-hand side, we obtain

$$p_{\alpha,s}(t+1)=\frac{\sum_{i} p_{\alpha,i}(t)\cdot r_{\alpha,i}(t)\cdot Q_{i\to s}(t)}{\bar{r}_{\alpha}(t)}.$$

To recover the replicator equation (A1), we define an effective replication rate that incorporates the effect of mutation:

$$r_{\alpha,s}^{(eff)}(t)=\frac{N_{\alpha,s}(t+1)}{N_{\alpha,s}(t)}=r_{\alpha,s}(t)\,Q_{s\to s}(t)+\sum_{i\neq s}\frac{N_{\alpha,i}(t)}{N_{\alpha,s}(t)}r_{\alpha,i}(t)\,Q_{i\to s}(t),$$

passing to frequencies,

$$r_{\alpha,s}^{eff}(t)=r_{\alpha,s}(t)Q_{s\to s}(t)+\sum_{i\neq s} r_{\alpha,i}(t)\left[Q_{i\to s}(t)/p_{\alpha,s}(t)\right]p_{\alpha,i}(t).$$

The term $r_{\alpha,s}^{eff}(t)$ meassures the effective fitness of class $s$.

Thus the replicator equation (A1) now becomes

$$p_{\alpha,s}(t+1)=\frac{p_{\alpha,s}(t)\cdot r_{\alpha,s}^{eff}(t)}{\bar{r}_{\alpha}^{eff}(t)}, \qquad \text{(A3)}$$

where

$$\bar{r}_\alpha^{eff}(t)=\sum_u p_{\alpha,u}(t)\cdot r_{\alpha,u}(t).$$

In this way, the effective replication rate measures the total change factor of the population, including both the descendants that remain in the same class and those that arrive from other classes.

### 10.2.5. Functional or Information Functor with Two Channels $J_r$ and $J_{rt}$

The information functor $\mathrm{Info}_\alpha$ acts on the replication morphisms $f_{\alpha,t}$, but its value is computed as the Jeffreys divergence between the frequency distributions induced by these morphisms through the population functo $P_\alpha$. The functor can be decomposed into two channels: the selective channel $\mathrm{Info}_\alpha^{(r)}$, which measures the information due exclusively to selection, and the total channel $\mathrm{Info}_\alpha^{(rt)}$, which also includes the effects of transmission.

#### *10.2.5.1. Basic notation*

$$p(t)=p_\alpha(t)=(p_{\alpha,0}(t),p_{\alpha,1}(t),\ldots),$$
$$\bar{r}_\alpha(t)=\sum_s p_{\alpha,s}(t)\cdot r_{\alpha,s}(t),$$
$$r'_{\alpha,s}(t)=r_{\alpha,s}(t)/\bar{r}_\alpha(t),$$
$$r'(t)=(r'_{\alpha,0}(t),r'_{\alpha,1}(t),\ldots).$$

#### *10.2.5.2. Update by Selection*

For each state $s$ at time $t$ ($s \leq t$), the state at $t$+1 is updated according to (A1), here written as

$$(p_r)_s(t+1)=r'_{\alpha,s}(t)\,p_{\alpha,s}(t).$$

#### *10.2.5.3. Update by Selection + Transmision*

Here, transmission includes the effects of mutation, migration, or class flow, absorbed into an effective replication rate. For each state $s$ at time $t$ ($s \leq t$), the state at $t$+1 is updated according to (A3).

#### *10.2.5.4. Informational Functionals*

*(a) Purely Selective Information*

$$Info_\alpha^{(r)}(f_{\alpha,t})=J(p_r(t+1),p(t))=\sum_s((p_r)_s(t+1)-p_{\alpha,s}(t))\ln\frac{(p_r)_s(t+1)}{p_{\alpha,s}(t)}.$$

*(b) Total Information (Selection + Transmission)*

$$Info_{\alpha}^{(rt)}(f_{\alpha,t}) = J(p(t+1), p(t)) = \sum_s (p_{\alpha,s}(t+1) - p_{\alpha,s}(t)) \ln \frac{p_{\alpha,s}(t+1)}{p_{\alpha,s}(t)},$$

where $J \geq 0$ and $J = 0$ iff the distributions at $t$ and $t$+1 are identical.

*(c) Transmissional (Non-Selective) Component*

For each step $t$, we obtain the partition:

$$Info_{\alpha}^{(rt)} = Info_{\alpha}^{(r)} + T_{trans}((\vec{p}), (\vec{r}), Q),$$

where

$$T_{trans}(f, r, Q) = \sum_s f_{\alpha,s}(t) [(\frac{r_{\alpha,s}^{eff}(t)}{\bar{r}_{\alpha}(t)} - 1) \ln \frac{r_{\alpha,s}^{eff}(t)}{\bar{r}_{\alpha}(t)} - (\frac{r_{\alpha,s}(t)}{\bar{r}_{\alpha}(t)} - 1) \ln \frac{r_{\alpha,s}(t)}{\bar{r}_{\alpha}(t)}]$$

is the component due to transition effects, as a function of frequencies, replication rates, and the transition matrix $Q$.

If $Q = I$ (identity) then $T_{trans} = 0$ and $J_{rt} = J_r$ the causes of change are then purely selective.

#### *10.2.5.5. Handling of Zeros*

In the case of appearance of a new class $p_{\alpha,s}(t)$=0, $p_{\alpha,s}(t+1)$>0 (e.g. $s = t$+1), the corresponding term in $J_{rt}$

$$(p_{\alpha,s}(t+1) - 0) \ln p_{\alpha,s}(t+1)/0$$

diverges. Conceptually, this can be interpreted as $J$ jumping to infinity when innovation (new information) arises.

In the case of disappearance of a class, $p_{\alpha,s}(t)$>0, $p_{\alpha,s}(t+1)$ =0, the object transitions to the empty set, conceptually not problematic, as it represents the loss of an informational channel.

In both cases, the problem arises only when computing $J$ explicitly. To handle this, we apply a regularization, assuming a smoothed version of the distributions such that the support is complete (no zeros). For both distributions, we assume

$$p^{\varepsilon} = (1 - \varepsilon) p + \varepsilon / K \quad (1 \gg \varepsilon > 0),$$

where $K$ is the number of classes.

Thus, for any *J*, we can write

$$J_{\varepsilon}(p(t+1),p(t))=\sum_{s}(p_s^{\varepsilon}(t+1)-p_s^{\varepsilon}(t))\ln\frac{p_s^{\varepsilon}(t+1)}{p_s^{\varepsilon}(t)}.$$

#### *10.2.5.6. Local Fold in $R_\alpha$*

$$Info_{\alpha}^{(rt)}(f_{\alpha,t})=J(p_{\alpha}(t+1),p_{\alpha}(t)),$$

$$Info_{\alpha}^{(rt)}(T)=\sum_{t=0}^{T-1}Info_{\alpha}^{(rt)}(f_{\alpha,t}),$$

measures the total information generated by selection and transmission within lineage α.

### 10.2.6. Reproduction *M*, Mating Fitness *m* and $J_{repr}$

#### *10.2.6.1. Reproductive Operator (Categorical Product)*

Reproduction is modeled as a binary operation (a reproductive bifunctor) between lineage objects

$$\mathrm{M}:\alpha_s(t)\times\beta_{s'}(t)\rightarrow(\alpha\beta)(t+1),$$

where M maps pairs of parental lineages to the lineage (or set of offspring lineages) they produce.

#### *10.2.6.2. Pure Replication*

In the case of pure replication, reproduction occurs within a single lineage and involves no pairing or combination between different entities. Two complementary types of morphisms can be distinguished:

**(a) Identity morphism (structural invariance)**

$$id_{\alpha(t)}:\alpha(t)\rightarrow\alpha(t).$$

This morphism is structural, not temporal. It represents the self-preservation or instantaneous maintenance of the informational handler in its current state, without invoking any passage of time. Biologically, it expresses the potential of the system to remain stable, preserving its organization and information content, within a given temporal frame.

**(b) Replication morphism (causal process)**

$$f_{\alpha,t}:\alpha_s(t)\rightarrow\alpha_{s'}(t+1).$$

This morphism represents the discrete temporal transition from step *t* to *t*+1. It may or may not involve a change in the internal state *s* depending whether replication occurs without mutation (*s*'=*s*), preserving the internal configuration or wheher replication occurs with mutation or variation ($s' \neq s$), producing a new subtype within the lineage. The morphism $f_{\alpha,t}$ therefore captures the causal process by which the system generates its next configuration, either identical or modified, maintaining lineage continuity while potentially introducing informational change.

### *10.2.6.3. Intra-Lineage Reproduction*

Within a lineage, reproduction

$$\mathrm{M}_{\alpha;s,s'}: \alpha_s(t)\times\alpha_{s'}(t) \rightarrow \alpha_{(s \boxplus s')}(t+1)$$

is represented by the discrete combination rule ⊞, that is, a function

$$\gamma: S\times S \rightarrow S,$$

that determines the type of offspring:

$$\alpha_s(t)\times\alpha_{s+1}(t) \rightarrow \alpha_{\gamma(s,s')}(t+1).$$

For example,

$$\alpha_s(t)\times\alpha_{s+1}(t) \rightarrow \alpha_{s+1}(t+1),$$

which represents that the lineage state $s$+1 dominates, or

$$\alpha_s(t)\times\alpha_{s'}(t) \rightarrow \alpha_{s''}(t+1)$$

represents the generation of a new subtype within lineage α.

### *10.2.6.4. Inter-Lineage Reproduction*

Between distinct lineages α and β, reproduction follows a cross-lineage generation rule

$$\mathrm{M}_{\alpha,\beta}: \alpha_s(t)\times\beta_{s'}(t) \rightarrow (\alpha\beta)_{\gamma(s,s')}(t+1),$$

where γ(s, s') generates combined states between lineages.

### *10.2.6.5. Mating frequencies*

Let $p_i(t)$ and $p_j(t)$ denote the frequencies of object $i$ in lineage α and object $j$ in lineage β at time $t$, respectively. The probability of random mating between them is:

$$q_{ij}(t)=p_i(t)\,p_j(t)$$

and the probability of mating influenced by mating fitness $m_{ij}$ is:

$$q'_{ij}(t)=q_{ij}(t)\,m_{ij}(t)/M(t),$$

where

$$M(t)=\sum_{i,j} q_{ij}(t)m_{ij}(t)$$

is the mean mating fitness.

After mating, the replication rates depend on the corresponding reproductive rule:

$$\phi_{\alpha,\beta}: r_\alpha \times r_\beta \rightarrow r_{\alpha\beta}^{post} \circ \hat{m}_{\alpha,\beta}.$$

In general form:

$$r_{ij}^{post}(t)=G(r_{\alpha,i}(t), r_{\beta,j}(t))$$

and, in specific cases:

$$G_L(x,y)=x, G_R(x,y)=y, G_{mean}(x,y)=(x+y)/2,$$

$$G_{geom}(x,y)=\sqrt{xy}, G_\lambda(x,y)=\lambda x+(1-\lambda),$$

$$G_{hibrid}(x,y)=I_{xy}(x+y)/2,$$

where $I_{xy} \in [0,+\infty)$ is a factor representing sexual isolation, where $I_{xy} = 0$ corresponds to complete isolation.

#### *10.2.6.6. Reproductive Fold (Jeffreys Divergence in Pair Space)*

The deviation from random mating is measured by the Jeffreys divergence in the space of pairs:

$$J_{repr}(q'(t), q(t))=\sum_{i,j}(q'_{ij}(t)-q_{ij}(t))\ln\frac{q'_{ij}(t)}{q_{ij}(t)}.$$

This measure can, in turn, be separated into additive components of male and female sexual selection and an assortative mating component [77].

#### *10.2.6.7. Decomposition by Lineage Relationship*

In this formulation, reproduction is decomposed into three informational channels, corresponding to the total reproductive channel $J_{repr}$, the intra-lineage channel $J_{repr,intra}$, and the inter-lineage channel $J_{repr,inter}$.

**Intra-lineage (within α ):**

$$J_{repr}^{intra,\alpha}=\sum_{i,j\in\alpha}(q'_{ij}-q_{ij})\ln\frac{q'_{ij}}{q_{ij}}.$$

***Inter-lineage (between α and β):***

$$J_{repr}^{inter,\alpha,\beta} = \sum_{i\in\alpha, j\in\beta} (q'_{ij} - q_{ij}) \ln \frac{q'_{ij}}{q_{ij}},$$

where the indices *i*, *j* range over the coexisting substates at step *t*.

In this partition of reproductive information, we distinguish the total, intra-lineage, and inter-lineage channels, each defined over its own normalized distribution. Since the normalizations differ, the corresponding divergences are not additive, but each preserves probability consistency and an informational interpretation within its respective domain.

### 10.2.7. Higher Level: Global Fold in *R*

The global fold integrates the total information generated by replication, transmission, and reproduction.

#### *10.2.7.1. Total information at t*

$$Info_{Total}(t) = J_{repr}(q'(t), q(t)) + \sum_{\alpha\in\Lambda} J_{rt}(f_\alpha(t+1), f_\alpha(t)).$$

#### *10.2.7.2. Global Information over T: Purely Selective Option*

$$Info_{Global}^{(r)}(T) = \sum_{\alpha,t} J_r(f_\alpha(t+1), f_\alpha(t)).$$

This expression measures the accumulated information due solely to selection across all lineages and discrete time steps.

#### *10.2.7.3. Global Information over T: Total Option (Selection + Transmission)*

$$Info_{Global}^{(rt)}(T) = \sum_{\alpha,t} J_{rt}(f_\alpha(t+1), f_\alpha(t)).$$

This includes selection plus the transmission effects (mutation, recombination, migration) accumulated through time and across all lineages.

#### *10.2.7.4. Explicit Decomposition of the Transmission Channel*

$$Info_{Global}^{rt} = Info_{Global}^{r} + \sum_{\alpha,t} T_{Trans,\alpha}(t),$$

$$\Delta\, Info_{trans}^{Global}(T) = Info_{Global}^{(rt)}(T) - Info_{Global}^{(r)}(T) = \sum_{\alpha\in\Lambda} \Delta\, Info_{Trans}(\alpha, T)$$

represents the total information associated with non-selective transmission effects.

### 10.2.7.5. Total Global Information in T

$$Info_{Global}^{Total}(T)=\sum_{t=0}^{T-1} Info_{Total}(t),$$

which captures the complete informational content accumulated by the system due to selection, transmission, and reproduction over the interval $[0,T)$.

## 10.3. MULTICOMPONENT MICROBIOME DECOMPOSITION AND THE GENERALIZED INFORMATION PARTITION

In many biological scenarios, the microbiome associated with a host is not adequately represented as a single lineage, but rather as a structured assemblage of multiple microbial components (e.g., phyla, functional guilds, or ASV-defined sublineages). To capture this structure, we extend the holobiont model by decomposing the microbiome into $K$ components, each with its own internal dynamics. A holobiont state is therefore indexed by a host type $i$ and a vector of microbiome types $(j_1,\ldots,j_K)$, where $j_k$ denotes the state of the $k$-th microbiome component.

Let the joint pre-selection frequency be

$$q_{i,j_1,\ldots,j_K}(t).$$

### 10.3.1. Fitness decomposition

We assume a multiplicative decomposition of absolute holobiont fitness

$$W_{i,j_1,\ldots,j_K}(t)=W_H(i,t)\cdot\prod_{k=1}^{K} W_{M_k}(j_k,t)\cdot\prod_{k=1}^{K} a_{i\,j_k}^{(k)}(t),$$

where

$W_H(i,t)$ is the host contribution,

$W_{M_k}(j_k,t)$ is the contribution of microbiome component $k$ and

$a_{i\,j_k}^{(k)}(t)$ is the host–component-$k$ association factor.

Under selection, the next-generation frequences follow the standard update:

$$q'_{i,j_1,\ldots,j_K}(t)=q_{i,j_1,\ldots,j_K}(t)\cdot\frac{W_{i,j_1,\ldots,j_K}(t)}{\bar{W}(t)}.$$

### 10.3.2. Relative log-fitness and informational partition

Define the relative log-fitness

$$Z_{i,j_1,\ldots,j_K}(t) := \log \frac{W_{i,j_1,\ldots,j_K}(t)}{\bar{W}(t)}.$$

Because relative log-fitness differs from absolute log-fitness by a constant

$$-\log(\bar{W})$$

and since

$$\sum_{i,j_1,\ldots,j_K} (q' - q) = 0,$$

the informational change due to selection is

$$J = \sum_{i,j_1,\ldots,j_K} (q'_{i,j_1,\ldots,j_K} - q_{i,j_1,\ldots,j_K}) Z_{i,j_1,\ldots,j_K}.$$

Using the multiplicative structure of the fitness, Z decomposes additively as:

$$Z_{i,j_1,\ldots,j_K} = Z_H(i) + \sum_{k=1}^{K} Z_{M_k}(j_k) + \sum_{k=1}^{K} Z_{assoc,k}(i, j_k),$$

where

$$Z_H(i) := \log W_H(i),$$

$$Z_{M_k}(j_k) := \log W_{M_k}(j_k),$$

$$Z_{assoc,k}(i, j_k) := \log a^{(k)}_{i j_k}.$$

By linearity this yields the multicomponent informational partition

$$J = J_H + \sum_{k=1}^{K} J_{M_k} + I_{assoc},$$

where

$$J_H = \sum_i (p'_i - p_i) Z_H(i),$$

$$J_{M_k} = \sum_{j_K} (m'_{k,j_K} - m_{k,j_K}) Z_{M_k}(j_k),$$

and

$$I_{assoc} = \sum_{i,j_1,\ldots,j_K} (q'_{i,j_1,\ldots,j_K} - q_{i,j_1,\ldots,j_K}) \sum_{k=1}^{K} Z_{assoc,k}(i, j_k).$$

These expressions are obtained from the joint form by summing over the microbiome (for $J_H$) or over all other indices (for $J_{Mk}$), using the fact that host marginals are given by

$$p_i=\sum_{j_1,\dots,j_K} q_{i,j_1,\dots,j_K};\;\; p'_i=\sum_{j_1,\dots,j_K} q'_{i,j_1,\dots,j_K}$$

and microbiome marginals

$$m_{k,j_k}=\sum_{i,j_1,\dots,j_{k-1},j_{k+1},\dots,j_K} q_{i,j_1,\dots,j_k,\dots,j_K};\;\; m'_{k,j_k}=\sum_{i,j_1,\dots,j_{k-1},j_{k+1},\dots,j_K} q'_{i,j_1,\dots,j_k,\dots,j_K},$$

where the summation runs over all indices except $j_k$.

Thus, the total information divides into a host component, a sum of microbiome components, and a multicomponent association term.

### 10.3.3. Partition of the multicomponent association term

To further decompose the association term into a pure association and an interaction component, we follow the approach used in Carvajal-Rodríguez [78] and in Section 7.4.4.

Let define the independence reference:

$$r_{i,j_1,\dots,j_K}:=p'_i\cdot\prod_{k=1}^{K} m'_{k,j_k},$$

that is, the independent distribution matching the post-selection marginals of $q'$.

As usual:

$$q'-q=(q'-r)+(r-q)$$

and

$$\log\frac{q'}{q}=\log\frac{q'}{r}+\log\frac{r}{q}.$$

Expanding the Jeffreys divergence

$$J=\sum(q'-q)\log\frac{q'}{q}$$

with these identities gives:

$$J=T_1+T_2+T_3+T_4,$$

where

$$T_1 = \sum (r - q) \log \frac{r}{q},$$

$$T_2 = \sum (r - q) \log \frac{q'}{r},$$

$$T_3 = \sum (q' - r) \log \frac{r}{q},$$

$$T_4 = \sum (q' - r) \log \frac{q'}{r}.$$

As in the two-component case, the term $T_3 = 0$ because $q'$ and $r$ share the same marginal distributions.

Therefore,

$$J = (J_H + \sum_{k=1}^{K} J_{M_k}) + J_{assoc} + E_{multi}, \qquad \text{(A4)}$$

where

$$T_1 = J_H + \sum_{k=1}^{K} J_{M_k},$$

$$J_{assoc} := T_4 = \sum (q' - r) \log \frac{q'}{r},$$

and

$$E_{multi} := T_2 = \sum (r - q) \log \frac{q'}{r}.$$

Finally:

$$I_{assoc} = J_{assoc} + E_{multi}.$$

Here, $J_{assoc}$ generalizes the "choice" component of the two-dimensional holobiont case and plays the role analogous to the $J_{PSI}$ term in mating models, in the sense that it quantifies the information contained in the full host-multicomponent microbiome association pattern, once the marginal changes have been accounted for. The term $E_{multi}$ is the interaction residual; it captures the non-additive information arising from the simultaneous presence of marginal and joint-structure changes, and constitutes the natural extension of both the $E$ term [77,78] and the holobiont interaction component $E_{holo}$.

Equation (A4) reduces to the two-component holobiont model when $K = 1$, and to the mating-based partitions when the two interacting components correspond to the sexes rather than to host and microbiome.

**Acknowledgments:** This work is dedicated to my sister Paci, a source of innocence, courage, and love. I would like to thank two anonymous reviewers for their helpful comments and suggestions.

**Funding:** This was funded by Xunta de Galicia (ED431C 2024/22), Ministerio de Ciencia e Innovación (PID2022-137935NB-I00) and Centro singular de investigación de Galicia accreditation 2024-2027 (ED431G 2023/07) and "ERDF A way of making Europe".

**Data Availability Statement:** No data were used in this study.

**Conflicts of interest:** The authors declare no conflicts of interest.

## REFERENCES

1. Margulis, L.; Sagan, D. *What Is Life?*; University of California Press, 2000; ISBN 978-0-520-22021-8.
2. McKay, C.P. What Is Life—and How Do We Search for It in Other Worlds? *PLOS Biology* **2004**, *2*, e302, doi:10.1371/journal.pbio.0020302.
3. Gómez-Márquez, J. What Is Life? *Mol Biol Rep* **2021**, *48*, 6223–6230, doi:10.1007/s11033-021-06594-5.
4. Mariscal, C. Life. In *The Stanford Encyclopedia of Philosophy*; Zalta, E.N., Ed.; Metaphysics Research Lab, Stanford University, 2021.
5. Skene, K.R. Systems Theory, Thermodynamics and Life: Integrated Thinking across Ecology, Organization and Biological Evolution. *Biosystems* **2024**, *236*, 105123, doi:10.1016/j.biosystems.2024.105123.
6. Lian, C.; Narimatsu, M.; Nara, K.; Hogetsu, T. Tricholoma Matsutake in a Natural Pinus Densiflora Forest: Correspondence between above- and below-Ground Genets, Association with Multiple Host Trees and Alteration of Existing Ectomycorrhizal Communities. *New Phytol* **2006**, *171*, 825–836, doi:10.1111/j.1469-8137.2006.01801.x.
7. Beiler, K.J.; Simard, S.W.; Lemay, V.; Durall, D.M. Vertical Partitioning between Sister Species of Rhizopogon Fungi on Mesic and Xeric Sites in an Interior Douglas-Fir Forest. *Mol Ecol* **2012**, *21*, 6163–6174, doi:10.1111/mec.12076.
8. Steffen, W.; Richardson, K.; Rockström, J.; Schellnhuber, H.J.; Dube, O.P.; Dutreuil, S.; Lenton, T.M.; Lubchenco, J. The Emergence and Evolution of Earth System Science. *Nat Rev Earth Environ* **2020**, *1*, 54–63, doi:10.1038/s43017-019-0005-6.
9. Sowinski, D.R.; Ghoshal, G.; Frank, A. Exo-Daisy World: Revisiting Gaia Theory through an Informational Architecture Perspective. *Planet. Sci. J.* **2025**, *6*, 176, doi:10.3847/PSJ/ade310.
10. Mossio, M.; Montévil, M.; Longo, G. Theoretical Principles for Biology: Organization. *Progress in Biophysics and Molecular Biology* **2016**, *122*, 24–35, doi:10.1016/j.pbiomolbio.2016.07.005.
11. Mossio, M. Introduction: Organization as a Scientific Blind Spot. In *Organization in Biology*; Mossio, M., Ed.; Springer International Publishing: Cham, 2024; pp. 1–22 ISBN 978-3-031-38968-9.

12. Bonduriansky, R.; Day, T. *Extended Heredity*; Princeton University Press, 2018; ISBN 978-0-691-15767-2.
13. Futuyma, D. Russell Bonduriansky and Troy Day. Extended Heredity: A New Understanding of Inheritance and Evolution. *Evolutionary Studies in Imaginative Culture* **2019**, *3*, 115–118, doi:10.26613/esic.3.1.130.
14. Dodd, M.S.; Papineau, D.; Grenne, T.; Slack, J.F.; Rittner, M.; Pirajno, F.; O'Neil, J.; Little, C.T.S. Evidence for Early Life in Earth's Oldest Hydrothermal Vent Precipitates. *Nature* **2017**, *543*, 60–64, doi:10.1038/nature21377.
15. Westall, F.; Brack, A.; Fairén, A.G.; Schulte, M.D. Setting the Geological Scene for the Origin of Life and Continuing Open Questions about Its Emergence. *Front Astron Space Sci* **2023**, *9*, 1095701, doi:10.3389/fspas.2022.1095701.
16. Sterelny, K.; Smith, K.C.; Dickison, M. The Extended Replicator. *Biol Philos* **1996**, *11*, 377–403, doi:10.1007/BF00128788.
17. Griffiths, P.E.; Gray, R.D. Replicator II – Judgement Day. *Biology & Philosophy* **1997**, *12*, 471–492, doi:10.1023/A:1006551516090.
18. Brenner, S. Sequences and Consequences. *Philosophical Transactions of the Royal Society B: Biological Sciences* **2010**, doi:10.1098/rstb.2009.0221.
19. Mazzocchi, F. Complexity and the Reductionism–Holism Debate in Systems Biology. *Wiley Interdisciplinary Reviews: Systems Biology and Medicine* **2012**, *4*, 413–427.
20. Nicholson, D.J. Is the Cell Really a Machine? *Journal of Theoretical Biology* **2019**, *477*, 108–126, doi:10.1016/j.jtbi.2019.06.002.
21. Hancock, Z.B.; Lehmberg, E.S.; Bradburd, G.S. Neo-Darwinism Still Haunts Evolutionary Theory: A Modern Perspective on Charlesworth, Lande, and Slatkin (1982). *Evol* **2021**, *75*, 1244–1255, doi:10.1111/evo.14268.
22. DiFrisco, J.; Gawne, R. Biological Agency: A Concept without a Research Program. *J Evol Biol* **2025**, *38*, 143–156, doi:10.1093/jeb/voae153.
23. Mazzocchi, F. An Investigation Into the Notion of Complex Systems. *Found Sci* **2025**, doi:10.1007/s10699-025-09975-2.
24. Noble, D.; Noble, R. How the Central Dogma and the Theory of Selfish Genes Misled Evolutionary and Medical Sciences in Understanding Multi-Factorial Diseases. *Evol Biol* **2025**, *52*, 138–148, doi:10.1007/s11692-025-09651-0.
25. Gilbert, W. Origin of Life: The RNA World. *Nature* **1986**, *319*, 618–618, doi:10.1038/319618a0.
26. Bernhardt, H.S. The RNA World Hypothesis: The Worst Theory of the Early Evolution of Life (except for All the Others)a. *Biol Direct* **2012**, *7*, 23, doi:10.1186/1745-6150-7-23.
27. Newman, S.A.; Benítez, M.; Bhat, R.; Glimm, T.; Kumar, K.V.; Nanjundiah, V.; Nicholson, D.J.; Sarkar, S. Agency in the Evolutionary Transition to Multicellularity. *The Quarterly Review of Biology* **2025**, *100*, 83–118, doi:10.1086/735964.
28. Shirt-Ediss, B.; Ferrero-Fernández, A.; De Martino, D.; Bich, L.; Moreno, A.; Ruiz-Mirazo, K. Modelling the Prebiotic Origins of Regulation and Agency in Evolving Protocell Ecologies. *Philosophical Transactions of the Royal Society B: Biological Sciences* **2025**, *380*, 20240287, doi:10.1098/rstb.2024.0287.
29. Moreno, Á.; Peretó, J. Metazoans: An Explosion of Agency. In *An Evolutionary Story of Agency: How Life Evolved to Act on its Own*; Moreno, Á., Peretó, J., Eds.; Springer Nature Switzerland: Cham, 2026; pp. 77–97 ISBN 978-3-032-05644-3.
30. Payseur, B.A.; Rieseberg, L.H. A Genomic Perspective on Hybridization and Speciation. *Molecular Ecology* **2016**, *25*, 2337–2360, doi:10.1111/mec.13557.
31. Johnson, N.A. Speciation: Genomic Sequence Data and the Biogeography of Speciation. *Natl Sci Rev* **2022**, *9*, doi:10.1093/nsr/nwac294.
32. Bock, D.G.; Cai, Z.; Elphinstone, C.; González-Segovia, E.; Hirabayashi, K.; Huang, K.; Keais, G.L.; Kim, A.; Owens, G.L.; Rieseberg, L.H. Genomics of Plant Speciation. *Plant Communications* **2023**, *4*, 100599, doi:10.1016/j.xplc.2023.100599.

33. Benítez-Benítez, C.; Mohan, A.V.; Sánchez-Villegas, R.; Gómez-Ramos, I.; Valdés-Florido, A.; Lucek, K.; Slovák, M.; Kolář, F.; Leitch, I.J.; Luceño, M.; et al. Bridging Micro and Macroevolution: Insights from Chromosomal Dynamics in Plants. *Front. Plant Sci.* **2025**, *16*, doi:10.3389/fpls.2025.1606450.

34. Moya, A.; Peretó, J.G. *Simbiosis: seres que evolucionan juntos*; Síntesis: Madrid, 2011; ISBN 978-84-9756-734-3.

35. Batstone, R.T. Genomes within Genomes: Nested Symbiosis and Its Implications for Plant Evolution. *New Phytologist* **2022**, *234*, 28–34, doi:10.1111/nph.17847.

36. Archibald, J.M. Symbiotic Revolutions at the Interface of Genomics and Microbiology. *PLOS Biology* **2024**, *22*, e3002581, doi:10.1371/journal.pbio.3002581.

37. Richards, T.A.; Moran, N.A. Symbiosis: In Search of a Deeper Understanding. *PLOS Biology* **2024**, *22*, e3002595, doi:10.1371/journal.pbio.3002595.

38. Wang, R.; Meng, Q.; Wang, X.; Xiao, Y.; Sun, R.; Zhang, Z.; Fu, Y.; Di Giuseppe, G.; Liang, A. Comparative Genomic Analysis of Symbiotic and Free-Living Fluviibacter Phosphoraccumulans Strains Provides Insights into the Evolutionary Origins of Obligate Euplotes–Bacterial Endosymbioses. *Applied and Environmental Microbiology* **2024**, *90*, e01900-23, doi:10.1128/aem.01900-23.

39. Miller, M.B.; Bassler, B.L. Quorum Sensing in Bacteria. *Annual Review of Microbiology* **2001**, *55*, 165–199, doi:10.1146/annurev.micro.55.1.165.

40. Ruan, Q.; Geng, S.; Yu, J.; Lu, L.; Liu, Y.; Chen, J.; Liao, Q.; Guo, R. Microbial Quorum Sensing: Mechanisms, Applications, and Challenges. *Biotechnology Advances* **2026**, *86*, 108733, doi:10.1016/j.biotechadv.2025.108733.

41. Dawkins, R. *The Selfish Gene*; Oxford university press, 2016; ISBN 0-19-878860-6.

42. Yockey, H.P. *Information Theory*, *Evolution*, *and the Origin of Life*; Cambridge University Press: New York, 2005; ISBN 0-521-80293-8.

43. Hansen, T.F. Fitness in Evolutionary Biology 2018.

44. Otsuka, J. The Role of Mathematics in Evolutionary Theory. *Elements in the Philosophy of Biology* **2019**, doi:10.1017/9781108672115.

45. Lewontin, R.C. The Units of Selection. *Annual Review of Ecology and Systematics* **1970**, *1*, 1–18.

46. Hull, D.L. Individuality and Selection. *Annual Review of Ecology*, *Evolution*, *and Systematics* **1980**, *11*, 311–332, doi:10.1146/annurev.es.11.110180.001523.

47. Okasha, S. *Evolution and the Levels of Selection*; Oxford : Oxford ; New York, 2008; ISBN 978-0-19-955671-7.

48. Suárez, J.; Lloyd, E.A. *Units of Selection*; Elements in the Philosophy of Biology; Cambridge University Press: Cambridge, 2023; ISBN 978-1-00-927641-2.

49. Marín, C. Three Types of Units of Selection. *Evolution* **2024**, *78*, 579–586, doi:10.1093/evolut/qpad234.

50. Dawkins, R. *Richard Dawkins The Extended Phenotype The Long Reach Of The Gene*; Oxford University Press, 1999; ISBN  978-0192880512.

51. Margalef, R. Information and Uncertainty in Living Systems, a View from Ecology. *Biosystems* **1996**, *38*, 141–146, doi:10.1016/0303-2647(95)01584-1.

52. Davies, P., C.W. and Walker, Sara Imari The Hidden Simplicity of Biology. *Reports on Progress in Physics* **2016**, *79*, 102601.

53. Davies, P. *The Demon in the Machine: How Hidden Webs of Information Are Solving the Mystery of Life*; University of Chicago Press: Chicago, IL, 2019; ISBN 978-0-226-66970-0.

54. Adami, C. *The Evolution of Biological Information: How Evolution Creates Complexity*, *from Viruses to Brains*; Princeton, 2024; ISBN 978-0-691-24114-2.

55. Paredes, O.; Farfán-Ugalde, E.; Gómez-Márquez, C.; Borrayo, E.; Mendizabal, A.P.; Morales, J.A. The Calculus of Codes - From Entropy, Complexity, and Information to Life. *Biosystems* **2024**, *236*, 105099, doi:10.1016/j.biosystems.2023.105099.

56. Fisher, R.A. *The Genetical Theory of Natural Selection*; Oxford University Press: Oxford, 1930;

57. Wright, S. Evolution in Mendelian Populations. *Genetics* **1931**, *16*, 97–159.

58. Haldane, J.B.S. *The Causes of Evolution*; Longmans, Green & Co.: London, 1932;

59. Dobzhansky, T.D. *Genetics and the Origin of Species: Columbia Classics Edition*; Columbia University Press, 1982; p. 364 Pages; ISBN 978-0-231-05475-1.

60. Lewontin, R.C. *The Genetic Basis of Evolutionary Change*; Columbia University Press: New York, 1974;

61. Oyama, S. *The Ontogeny of Information: Developmental Systems and Evolution*; Duke University Press, 2000; ISBN 978-0-8223-2431-7.

62. Jablonka, E.; Lamb, M.J. *Evolution in Four Dimensions: Genetic, Epigenetic, Behavioral, and Symbolic Variation in the History of Life*; Life and Mind: Philosophical Issues in Biology and Psychology; MIT Press: Cambridge, MA, USA, 2006; ISBN 978-0-262-60069-9.

63. Noble, D. A Theory of Biological Relativity: No Privileged Level of Causation. *Interface Focus* **2011**, *2*, 55–64, doi:10.1098/rsfs.2011.0067.

64. Marín, C.; Wade, M.J. Bring Back the Phenotype. *New Phytologist* **2025**, *246*, 2440–2445, doi:10.1111/nph.70138.

65. Adami, C. Information Theory in Molecular Biology. *Physics of Life Reviews* **2004**, *1*, 3–22, doi:10.1016/j.plrev.2004.01.002.

66. Koonin, E.V. The Meaning of Biological Information. *Philos Trans A Math Phys Eng Sci* **2016**, *374*, doi:10.1098/rsta.2015.0065.

67. Azpeitia, E.; Balanzario, E.P.; Wagner, A. Signaling Pathways Have an Inherent Need for Noise to Acquire Information. *BMC Bioinformatics* **2020**, *21*, 462, doi:10.1186/s12859-020-03778-x.

68. Nurse, P. Life, Logic and Information. *Nature* **2008**, *454*, 424–426, doi:10.1038/454424a.

69. Cartwright, J.H.E.; Čejková, J.; Fimmel, E.; Giannerini, S.; Gonzalez, D.L.; Goracci, G.; Grácio, C.; Houwing-Duistermaat, J.; Matić, D.; Mišić, N.; et al. Information, Coding, and Biological Function: The Dynamics of Life. *Artificial Life* **2024**, *30*, 16–27, doi:10.1162/artl_a_00432.

70. Garcia, M.; Theunissen, F.; Sèbe, F.; Clavel, J.; Ravignani, A.; Marin-Cudraz, T.; Fuchs, J.; Mathevon, N. Evolution of Communication Signals and Information during Species Radiation. *Nat Commun* **2020**, *11*, 4970, doi:10.1038/s41467-020-18772-3.

71. Kimura, M. Natural Selection as the Process of Accumulating Genetic Information in Adaptive Evolution. *Genetics Research* **1961**, *2*, 127–140, doi:10.1017/S0016672300000616.

72. Wagner, A. From Bit to It: How a Complex Metabolic Network Transforms Information into Living Matter. *BMC Systems Biology* **2007**, *1*, 33, doi:10.1186/1752-0509-1-33.

73. Adami, C. The Use of Information Theory in Evolutionary Biology. *Annals of the New York Academy of Sciences* **2012**, *1256*, 49–65.

74. Baalen, M. van Biological Information: Why We Need a Good Measure and the Challenges Ahead. *Interface Focus* **2013**, doi:10.1098/rsfs.2013.0030.

75. Wagner, A. Information Theory, Evolutionary Innovations and Evolvability. *Philosophical Transactions of the Royal Society B: Biological Sciences* **2017**, *372*, 20160416, doi:10.1098/rstb.2016.0416.

76. Frank, S.A. Natural Selection. V. How to Read the Fundamental Equations of Evolutionary Change in Terms of Information Theory. *J Evol Biol* **2012**, *25*, 2377–2396.

77. Carvajal-Rodríguez, A. Non-Random Mating and Information Theory. *Theoretical Population Biology* **2018**, *120*, 103–113, doi:10.1016/j.tpb.2018.01.003.

78. Carvajal-Rodríguez, A. Unifying Quantification Methods for Sexual Selection and Assortative Mating Using Information Theory. *Theoretical Population Biology* **2024**, *158*, 206–215, doi:10.1016/j.tpb.2024.06.007.

79. Hledík, M.; Barton, N.; Tkačik, G. Accumulation and Maintenance of Information in Evolution. *Proceedings of the National Academy of Sciences* **2022**, *119*, e2123152119, doi:10.1073/pnas.2123152119.

80. Smith, E. Beyond Fitness: The Information Imparted in Population States by Selection throughout Lifecycles. *Theoretical Population Biology* **2024**, *157*, 86–117, doi:10.1016/j.tpb.2024.04.003.

81. Kullback, S. *Information Theory and Statistics*; New edition.; Dover Publications: Mineola, N.Y, 1997; ISBN 978-0-486-69684-3.

82. Ay, N.; Jost, J.; Lê, H.V.; Schwachhöfer, L. *Information Geometry*; Ergebnisse der Mathematik und ihrer Grenzgebiete 34; Springer International Publishing: Cham, 2017; Vol. 64; ISBN 978-3-319-56477-7.

83. Carvajal-Rodríguez, A. On Non-Random Mating, Adaptive Evolution, and Information Theory. *Biology* **2024**, *13*, 970, doi:10.3390/biology13120970.

84. Margalef, R. *Perspectives in Ecological Theory*; University of Chicago Press, 1968; ISBN 978-0-226-50505-3.

85. Sherwin, W.B. Entropy, or Information, Unifies Ecology and Evolution and Beyond. *Entropy* **2018**, *20*, 727, doi:10.3390/e20100727.

86. Konopiński, M.K. Shannon Diversity Index: A Call to Replace the Original Shannon's Formula with Unbiased Estimator in the Population Genetics Studies. *PeerJ* **2020**, *8*, e9391, doi:10.7717/peerj.9391.

87. Pardo, L. Statistical Inference Based on Divergence Measures.; Chapman and Hall/CRC, November 12 2018.

88. Carvajal-Rodríguez, A. A Generalization of the Informational View of Non-Random Mating: Models with Variable Population Frequencies. *Theoretical Population Biology* **2019**, *125*, 67–74, doi:10.1016/j.tpb.2018.12.004.

89. Carvajal-Rodríguez, A. Multi-Model Inference of Non-Random Mating from an Information Theoretic Approach. *Theoretical Population Biology* **2020**, *131*, 38–53, doi:10.1016/j.tpb.2019.11.002.

90. Frank, S.A. Two Principles of Success. *Proceedings of the National Academy of Sciences* **2024**, *121*, e2417410121, doi:10.1073/pnas.2417410121.

91. Frank, S.A. Natural Selection at Multiple Scales. *Evolution* **2025**, *79*, 1166–1184, doi:10.1093/evolut/qpaf037.

92. Hardin, G. The Tragedy of the Commons. *Science* **1968**, *162*, 1243–1248, doi:10.1126/science.162.3859.1243.

93. Edelaar, P.; Bolnick, D.I. Appreciating the Multiple Processes Increasing Individual or Population Fitness. *Trends in Ecology & Evolution* **2019**, *34*, 435–446, doi:10.1016/j.tree.2019.02.001.

94. Bosch, T.C.G.; Wigley, M.; Colomina, B.; Bohannan, B.; Meggers, F.; Amato, K.R.; Azad, M.B.; Blaser, M.J.; Brown, K.; Dominguez-Bello, M.G.; et al. The Potential Importance of the Built-Environment Microbiome and Its Impact on Human Health. *Proceedings of the National Academy of Sciences* **2024**, *121*, e2313971121, doi:10.1073/pnas.2313971121.

95. Margulis, L. Genetic and Evolutionary Consequences of Symbiosis. *Experimental Parasitology* **1976**, *39*, 277–349, doi:10.1016/0014-4894(76)90127-2.

96. Margulis, L.; Bermudes, D. Symbiosis as a Mechanism of Evolution: Status of Cell Symbiosis Theory. *Symbiosis (Philadelphia, PA)* **1985**, *1*, 101–123.

97. Margulis, L. *Symbiogenesis and Symbionticism*; Conference on Symbiosis as a Source of Evolutionary Innovation (1989 : : Bellagio, I., Ed.; MIT Press: Cambridge, MA, 1991;

98. Roughgarden, J. Holobiont Evolution: Mathematical Model with Vertical vs. Horizontal Microbiome Transmission. *Philosophy, Theory, and Practice in Biology* **2020**, *12*, doi:https://doi.org/10.3998/ptpbio.16039257.0012.002.

99. Zilber-Rosenberg, I.; Rosenberg, E. Role of Microorganisms in the Evolution of Animals and Plants: The Hologenome Theory of Evolution. *FEMS Microbiol Rev* **2008**, *32*, 723–735, doi:10.1111/j.1574-6976.2008.00123.x.

100. Theis, K.R.; Dheilly, N.M.; Klassen, J.L.; Brucker, R.M.; Baines, J.F.; Bosch, T.C.G.; Cryan, J.F.; Gilbert, S.F.; Goodnight, C.J.; Lloyd, E.A.; et al. Getting the Hologenome Concept Right: An Eco-Evolutionary Framework for Hosts and Their Microbiomes. *mSystems* **2016**, *1*, e00028-16, doi:10.1128/mSystems.00028-16.

101. Week, B.; Russell, S.L.; Schulenburg, H.; Bohannan, B.J.M.; Bruijning, M. Applying Evolutionary Theory to Understand Host–Microbiome Evolution. *Nat Ecol Evol* **2025**, 1–12, doi:10.1038/s41559-025-02846-w.

102. Rashevsky, N. Topology and Life: In Search of General Mathematical Principles in Biology and Sociology. *Bulletin of Mathematical Biophysics* **1954**, *16*, 317–348, doi:10.1007/BF02484495.

103. Rosen, R. *Life Itself: A Comprehensive Inquiry Into the Nature, Origin, and Fabrication of Life*; Columbia University Press, 1991; ISBN 978-0-231-07564-0.

104. Mac Lane, S. *Categories for the Working Mathematician*; Graduate texts in mathematics; Springer: Heidelberg, 1971; ISBN 978-3-540-90036-8.

105. Cheng, E. *The Joy of Abstraction: An Exploration of Math, Category Theory, and Life*; Cambridge University Press: Cambridge, 2022; ISBN 978-1-108-47722-2.

106. Leinster, T. Basic Category Theory 2025.

107. Lane, P.A. Robert Rosen's Relational Biology Theory and His Emphasis on Non-Algorithmic Approaches to Living Systems. *Mathematics* **2024**, *12*, 3529, doi:10.3390/math12223529.

108. Varela, F.G.; Maturana, H.R.; Uribe, R. Autopoiesis: The Organization of Living Systems, Its Characterization and a Model. *Biosystems* **1974**, *5*, 187–196, doi:10.1016/0303-2647(74)90031-8.

109. Maturana, H.; Varela, F.J. *De máquinas y seres vivos : autopoiesis : la organización de lo vivo*; 2005th ed.; Buenos Aires, 2005; ISBN 978-987-00-0386-1.

110. Mossio, M.; Moreno, A. Organisational Closure in Biological Organisms. *History and Philosophy of the Life Sciences* **2010**, *32*, 269.

111. Pattee, H.H. The Physics of Symbols: Bridging the Epistemic Cut. *Biosystems* **2001**, *60*, 5–21, doi:10.1016/S0303-2647(01)00104-6.

112. Walker, S.I.; Davies, P.C.W. The Algorithmic Origins of Life. *Journal of The Royal Society Interface* **2013**, *10*, 20120869, doi:10.1098/rsif.2012.0869.

113. Tuyéras, R. Category Theory for Genetics I: Mutations and Sequence Alignments. *Theory and Applications of Categories* **2018**, *33*, 1269–1317.

114. Tuyéras, R. Category Theory for Genetics II: Genotype, Phenotype and Haplotype 2023.

115. Baez, J.C.; Otter, N. Operads and Phylogenetic Trees. *Theory and Applications of Categories* **2017**, *32*, 1397–1453.

116. Mahadevan, S. Universal Imitation Games 2024.

117. Wu, Y. A Category of Genes 2023.

118. Rodrigo, A.G. Modelling the Evolution of Holobionts: An Incomplete Review. *New Zealand Journal of Zoology* **2025**, *52*, 87–102, doi:10.1080/03014223.2023.2255142.

119. Bordenstein, S.R.; THE HOLOBIONT BIOLOGY NETWORK The Disciplinary Matrix of Holobiont Biology. *Science* **2024**, *386*, 731–732, doi:10.1126/science.ado2152.

120. Dawkins, R. Replicator Selection and the Extended Phenotype. *Zeitschrift für Tierpsychologie* **1978**, *47*, 61–76, doi:10.1111/j.1439-0310.1978.tb01823.x.

121. Bürger, R. *The Mathematical Theory of Selection, Recombination, and Mutation*; Wiley series in mathematical and computational biology; John Wiley: Chichester, 2000; ISBN 0-471-98653-4.

122. Laland, K.N.; Uller, T.; Feldman, M.W.; Sterelny, K.; Müller, G.B.; Moczek, A.; Jablonka, E.; Odling-Smee, J. The Extended Evolutionary Synthesis: Its Structure, Assumptions and Predictions. *Proc Biol Sci* **2015**, *282*, 20151019, doi:10.1098/rspb.2015.1019.

123. Mossio, M.; Pontarotti, G. Conserving Functions across Generations: Heredity in Light of Biological Organization. *The British Journal for the Philosophy of Science* **2019**, *73*, 249–278, doi:10.1093/bjps/axz031.

124. Noble, D.; Noble, R. How Purposive Agency Became Banned from Evolutionary Biology. In *Evolution "On Purpose": Teleonomy in Living Systems*; Corning, P.A., Kauffman, S.A., Noble, D., Shapiro, J.A., Vane-Wright, R.I., Pross, A., Eds.; The MIT Press, 2023; p. 0 ISBN 978-0-262-37601-3.

125. Vega, F. The Cell as a Semiotic System That Realizes Closure to Efficient Causation: The Semiotic (M, R) System. *BioSystems* **2024**, *240*, 105226, doi:10.1016/j.biosystems.2024.105226.

126. Montévil, M.; Mossio, M. Biological Organisation as Closure of Constraints. *J Theor Biol* **2015**, *372*, 179–191, doi:10.1016/j.jtbi.2015.02.029.

127. Avery, J.S. *Information Theory and Evolution*; 3rd ed.; World Scientific, 2021; ISBN 9789811250361.

128. Barbieri, M. A General Model on the Origin of Biological Codes. *Biosystems* **2019**, *181*, 11–19, doi:10.1016/j.biosystems.2019.04.010.

129. Prinz, R. Nothing in Evolution Makes Sense except in the Light of Code Biology. *Biosystems* **2023**, *229*, 104907, doi:10.1016/j.biosystems.2023.104907.

130. Yurchenko, S.B. Is Information the Other Face of Causation in Biological Systems? *Biosystems* **2023**, *229*, 104925, doi:10.1016/j.biosystems.2023.104925.

131. García-Sancho, M. From Metaphor to Practices: The Introduction of “Information Engineers” into the First DNA Sequence Database. *Hist Philos Life Sci* **2011**, *33*, 71–104.

132. Anta, J.; Sánchez-Dorado, J. The Pursuitworthiness of Informational Biology. *Biol Philos* **2025**, *40*, 21, doi:10.1007/s10539-025-09996-y.

133. Griffiths, P.E. Genetic Information: A Metaphor In Search of a Theory. *Philosophy of Science* **2001**, *68*, 394–412, doi:10.1086/392891.

134. Frank, S. The Price Equation Program: Simple Invariances Unify Population Dynamics, Thermodynamics, Probability, Information and Inference. *Entropy* **2018**, *20*, 978, doi:10.3390/e20120978.

135. Frank, S.A. The Price Equation Reveals a Universal Force–Metric–Bias Law of Algorithmic Learning and Natural Selection. *Entropy* **2025**, *27*, 1129, doi:10.3390/e27111129.

136. Sherwin, W.B. Pan-Evo: The Evolution of Information and Biology’s Part in This. *Biology* **2024**, *13*, 507, doi:10.3390/biology13070507.